\documentclass[12pt,draftclsnofoot,onecolumn]{IEEEtran}

\usepackage{}
\usepackage{graphicx}
\usepackage{epstopdf}
\usepackage{amssymb}
\usepackage{amsfonts}
\usepackage{amsmath}
\usepackage{algorithm}
\usepackage{algorithmic}
\usepackage{subeqnarray}
\usepackage{cases}
\usepackage{makecell}
\usepackage{stfloats,color}
\usepackage{bm}
\usepackage{subfigure,amsmath,amssymb,cite}

\begin{document}

\title{Beamforming and Power Allocation for Double-RIS-aided Two-way Directional Modulation Network}
\author{Rongen Dong, Feng Shu, Ri-Qing Chen, Yongpeng Wu, Cunhua Pan, and Jiangzhou Wang,\emph{ Fellow, IEEE}

\thanks{This work was supported in part by the National Natural Science Foundation of China (Nos. 62071234, 62071289, and 61972093), the Hainan Major Projects (ZDKJ2021022),  the Scientific Research Fund Project of Hainan University under Grant KYQD(ZR)-21008 and KYQD(ZR)-21007, and the National Key R\&DProgram of China under Grant 2018YFB1801102.}
\thanks{Rongen Dong and Feng Shu are with the School of Information and Communication Engineering, Hainan University, Haikou, 570228, China.}
\thanks{Ri-Qing Chen is with the College of Computer and Information Science Digital Fujian Research Institute of Big Data for Agriculture and Forestry Fujian Agriculture and Forestry University, Fuzhou, China (riqing.chen@fafu. edu.cn).}
\thanks{Yongpeng Wu is with the Shanghai Key Laboratory of Navigation and Location Based Services, Shanghai Jiao Tong University, Minhang, Shanghai 200240, China (e-mail: yongpeng.wu2016@gmail.com).}
\thanks{Cunhua Pan is with the National Mobile Communications Research Laboratory, Southeast University, China. (Email: cpan@seu.edu.cn).}
\thanks{Jiangzhou Wang is with the School of Engineering, University of Kent, Canterbury CT2 7NT, U.K. Email: {j.z.wang}@kent.ac.uk.}

}


\maketitle

\begin{abstract}
To improve the  information exchange rate between Alice and Bob in traditional two-way directional modulation (TWDM) network,  a new double-reconfigurable intelligent surface (RIS)-aided TWDM network is proposed. To achieve the low-complexity  transmitter design,  two analytical precoders, one closed-form method of adjusting the RIS phase-shifting matrices, and semi-iterative power allocation (PA) strategy of maximizing secrecy sum rate (SSR) are proposed. First, the geometric parallelogram (GPG) criterion is employed to give the  phase-shifting matrices of RISs. Then, two precoders, called maximizing singular value (Max-SV) and  maximizing signal-to-leakage-noise ratio (Max-SLNR), are proposed to enhance the SSR. Evenly, the maximizing SSR PA with hybrid iterative closed-form (HICF) is further proposed to improve the SSR  and derived to be  one root of a sixth-order polynomial computed by: (1) the Newton-Raphson algorithm is repeated twice to reduce the order of the polynomial from six to four; (2) the remaining four feasible solutions can be directly obtained by the Ferrari's method.  Simulation results show that using the proposed Max-SV and Max-SLNR, the proposed GPG makes a significant SSR improvement over random phase and no RIS. Given GPG, the proposed Max-SV outperforms the proposed leakage for small-scale or medium-scale RIS. Particularly, the proposed HICF PA stragey shows about ten percent  performance gain over equal PA.
\end{abstract}
\section{Introduction}


Directional modulation (DM), as a key method of physical layer security (PLS), is attracting much attention from academia and industry due to its future great promising applications in civil and military \cite{Shiu2011Physical, Larsson2014Massive, Akdeniz2014Millimeter, Mukherjee2015Physical, Ai2019Physical, Chen2017A, Wu2018A, Zou2018Security, Babakhani2008Transmitter, Daly2009Directional, Daly2010Demonstration, Daly2010Beamsteering}. Its basic idea is as follows:  in line-of-propagation (LoP) channel, transmit beamforming and artificial noise (AN)  are two main ways to improve the secure performance. The former  uses the beamforming vector to enhance the confidential message along the desired direction while the latter is projected along the undesired direction  to severely degrade the performance at Eve.

In \cite{Hong2011Dual}, the authors proposed a dual-beam DM scheme, in which the in-phase and quadrature baseband signals were used to excite two different antennas.  In \cite{Wan2018Power}, a general power allocation (PA) strategy of maximizing secrecy rate (Max-SR), given the  null-space projection (NSP) beamforming method, were proposed for secure DM network. In \cite{Teng2021low}, the authors considered a scenario for DM network with a full-duplex (FD) malicious attacker, where three high-performance receive beamforming methods were proposed to alleviate the impact of the jamming signal on the desired user. In DM, the beamforming vector and AN projection matrix are intimately related to the desired and undesired directions, Alice should behave as a receiver to make direction of arrival (DOA) measurements before performing a beamforming operation. In \cite{Shu2018Low}, to achieve low-complexity and high-resolution DOA estimation for practical DM,  a fast root multiple signal classification hybrid analog-digital (HAD) phase alignment method of DOA in hybrid MIMO structure were proposed.  In \cite{Zhuang2020Machine}, using the probability density function of measured DOA of a desired user, an AN-aided robust HAD plus DM transmitter was presented, and a robust and secure physical-layer transmission was achieved. In fact, for DM, the direction angle is not always perfect, for the imperfect direction angle, a low-complexity robust synthesis method for secure DM was proposed to make an one-order improvement on bit error
rate performance compared to non-robust ones in \cite{Hu2016Robust}.

To address the secure risk of DM that Eve moves to the desired main-beam from Alice to Bob and may eavesdrop the confidential message (CM) due to  its property of only depending on angle dimension,  in \cite{Hu2017Artificial},  a random frequency diverse array (FDA)-based DM scheme of randomly allocating frequencies to transmit antennas was proposed to implement a two-dimensional secure transmission of depending on both angle and range. In \cite{Shu2018Secure}, combining the orthogonal frequency division multiplexing and DM, a new secure precise wireless transmission concept was proposed to make it easy to implement in practice by  replacing random frequency diverse with random subcarrier selection.  A FDA-based DM aided by AN was proposed in \cite{Qiu2019Multi}, the AN projection matrix was calculated to minimize the effect of AN on legitimate user in the cases of known and unknown Eve locations.  In \cite{Cheng2021Physical}, a single-point AN-aided FDA DM scheme was proposed, where the FDA was analyzed in three-dimensional (i.e., range, azimuth angle, and elevation angle), compared with the conventional zero-forcing and singular value decomposition methods, and this method reduces the memory consumption significantly.

With the rapid development of wireless networks, there is a strong demand for a wireless network with lower implementation cost and energy consumption. Reconfigurable intelligent surface (RIS), consisting of a large number of small and low-cost reconfigurable passive elements, will meet this demand\cite{Wu2019Intelligent, Pan2020Multicell, Tang2021Wireless}. Actually, RIS is a passive forwarding device, which is viewed as a low-cost and low-energy-consumption reflecting relay. An intelligent reflecting surface (IRS)-aided simultaneous wireless information and power transfer (SWIPT) for multiple-input single-output (MISO) system was presented in \cite{Shi2020Enhanced} to maximize the harvested  energy by jointly optimizing the transmit beamforming and IRS phase shift. In \cite{Shen2020Beamforming}, an IRS-aided FD communication system was established to maximize the sum rate of two-way transmissions.  Compared with the Arimoto-Blahut method, this method achieved a  faster convergence rate and lower computational complexity. An IRS-aided decode-and-forward relay network system was investigated with multiple antennas at relay station in \cite{Wang2021Beamforming}, three maximizing receive power methods were proposed to achieve a high rate. In  a double-IRS-aided multi-user system \cite{Zheng2021Double},  the maximizing the minimum signal-to-interference-plus-noise ratio of all users was proposed to jointly optimize  the (active) receiving beamforming of the base station and (passive) cooperative reflection beamforming of the two distributed IRSs. A double-IRS-assisted wireless system was proposed in \cite{Tian2021Cooperative}, using the particle swarm optimization algorithm, the transmit and passive beamforming vectors on the two IRSs were cooperatively optimized to maximize the received signal power.

To explore the security of RIS-assisted wireless system, in \cite{Guan2020Intelligent}, the authors analyzed whether AN is helpful to enhance PLS, and identified the most beneficial practical scenario for using AN. In \cite{Wang2020Intelligent}, the authors investigated the improved  security of an IRS-assisted MISO  system, the oblique manifold and Majorization-Minimization algorithms were proposed to jointly optimize the transmit beamforming at transmitter and phase shifts at IRS. In \cite{Hong2020Artificial}, the IRS was used to enhance the security performance in MIMO system in order to maximize secrecy rate (SR). Here,  the block coordinate descent algorithm was proposed to alternately update the transmit precoding, AN covariance, and IRS phase shifting matrix. An IRS-aided secure spatial modulation system was presented in \cite{Shu2021Beamforming},  and three IRS beamforming methods and two transmit power design methods were proposed to improve the SR. A robust transmission design for an IRS-aided secure system in the presence of transceiver hardware impairments was investigated in \cite{Zhou2021Secure}, and an alternate optimization method was proposed to maximize the SR.

To enhance the energy efficiency and overcome the limitation of only one confidential signal being transmitted to legitimate user in the traditional DM network, in \cite{Shu2021Enhanced}, with the help of an IRS, the DM system has implemented two parallel independent confidential bit stream (CBS) transmission from Alice to Bob, where the general alternating iterative (GAI) algorithm and  low-complexity NSP algorithm were proposed to maximize the SR. They showed that the proposed two-stream transmission  approximately doubles the SR of conventional DM system in terms of  SR.   In \cite{Lai2020Directional},  an IRS-aided DM with AN scheme was proposed to achieve an enhanced secure single-stream transmission, and its closed-form expression for SR  was derived.

Although two CBSs  in \cite{Shu2021Enhanced} were independently and concurrently transmitted from Alice to Bob with the aid of RIS, only one-way information was sent from Alice to Bob. In this paper, we  propose a completely distinct new network, i.e., a new kind of two-way DM network aided by RIS. In other words, Alice and Bob exchange their messages each other via two RISs at the same time, which will be shown to significantly improve the SR of the traditional two-way DM network without RISs in our paper.
The main contributions of this paper are summarized as follows:
\begin{enumerate}
\item To enhance the secrecy sum rate (SSR)  performance and energy efficiency in the traditional DM networks, a double-RIS-aided two-way DM system is established. Here, both Alice and Bob work in FD mode, and friendly multipaths between Alice and Bob are created and controlled by the two RISs.   To maximize the SSR of this system, the phase-shifting matrices of two RISs are firstly designed and optimized  by using the geometric parallelogram (GPG) criterion, i.e., each RIS phase-shifting matrix  is chosen  to be negative to the phase part of  a synthesis  vector of two channel vectors independently reflected from RIS to Alice and Bob. In the simulation, it is verified compared to random phase method, the proposed GPG method can make a substantial SSR enhancement.
\item Given that RIS phase-shifting matrix has been designed by GPG strategy, one transmit beamforming scheme, called maximizing singular value (Max-SV), is proposed. Here, the right singular corresponding the maximum singular-value is used as the beamforming vector of the CM while the AN beamforming vector is designed on the null-space of the remaining singular vectors. Additionally, the maximizing signal-to-leakage-noise ratio (Max-SLNR) is generalized to the double-RIS-aided two-way DM network.  At Eve, a zero-forcing (ZF)-based maximum ratio combiner (MRC) method is proposed to achieve a high-performance receive beamforming.  Simulation results show that the proposed Max-SV  and generalized leakage methods outperform random phase and no RIS in terms of  SSR.
\item To further improve SSR, a PA strategy of maximizing SSR is proposed, which is addressed by the hybrid iterative-closed form (HICF) algorithm. Here,  the optimal PA factor is shown to be one root of a sixth-order polynomial. The HICF method consists of two steps:  In the first step, the Newton-Raphson algorithm is repeated twice to obtain two candidate roots and  reduce the order of the polynomial from six to four, and the remaining four feasible solutions can be obtained by the Ferrari's method; secondly, the optimal root is obtained by maximizing the SSR  over the set of six candidate roots and boundary points. Moreover, the two-dimensional  exhaustive search (2D-ES) algorithm is presented as a performance benchmark. Simulation results show that  the proposed HICF  achieves about a $10\%$ performance gain over  equal PA (EPA).
\end{enumerate}

The remainder of this paper is organized as follows. Section II describes the system model and problem formulation of the double-RIS-aided two-way DM network.
In Section III, two transmit beamforming methods are presented.
One PA scheme for maximizing SSR is given in Section IV.
Numerical simulation results are presented in Section V. Finally, we draw our conclusions in Section VI.

$\textbf{Notations}$: throughout this paper, boldface lower case and upper case letters represent vectors and matrices, respectively. Signs $(\cdot)^T$, $(\cdot)^*$, $(\cdot)^H$, $(\cdot)^{-1}$, $(\cdot)^{\dagger}$, tr$(\cdot)$, and $\|\cdot\|$ denote the transpose operation, conjugate operation, conjugate transpose operation, inverse operation, pseudo inverse operation, trace operation, and 2-norm operation, respectively. The symbol $\mathbb{C}^{N\times N}$ is the space of $N\times N$ complex-valued matrix. The notation $\textbf{I}_N$ is the $N\times N$ identity matrix. The sign $\mathbb{E}\{\cdot\}$ represents the expectation operation.

\section{system model and problem formulation}
\begin{figure}[htbp] 
\centering
\includegraphics[width=0.48\textwidth]{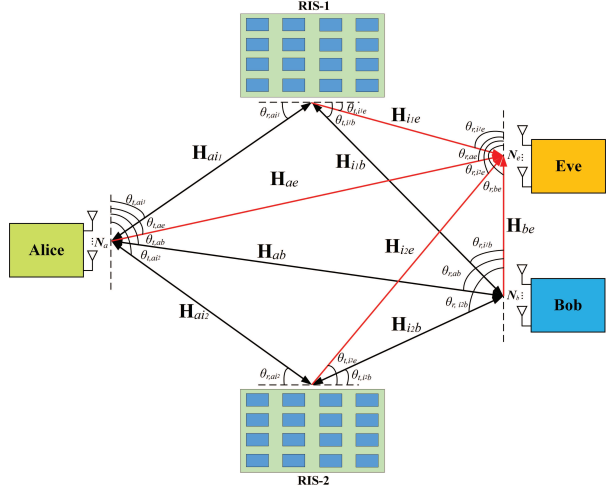}\\
\caption{System model diagram for double-RIS-aided two-way DM network.}\label{System-model}
\end{figure}

As shown in Fig.~\ref{System-model}, a double-RIS-aided two-way DM system is considered, where Alice is equipped with $N_a$ antennas, Bob is equipped with $N_{b}$ antennas, and an eavesdropper (Eve) is equipped with $N_e$ antennas, both of RIS-1 and RIS-2 are equipped with $M$ low-cost passive reflecting elements. The RIS reflects signal only one time slot. Both of Alice and Bob work in FD model.  For convenience of analysis and derivation below, it is assumed the self-interference is completely removed by the transmitters at Alice and Bob. The channels from Alice to RIS-1, Alice to RIS-2, Alice to Eve, Alice to Bob, RIS-1 to Eve, RIS-2 to Eve, Bob to Eve, RIS-1 to Bob, and RIS-2 to Bob are the line-of-propagation channel.

The transmit signal from Alice is
\begin{align}\label{1}
\textbf{S}_a&=\sqrt{\beta_1P_a} \textbf{v}_{at}x_1+
\sqrt{(1-\beta_1)P_a}\textbf{w}_{a},
\end{align}
where $P_a$ denotes the total transmit power, $\beta_1$ and $(1-\beta_1)$ represent the PA parameters of CM and AN, respectively. $\textbf{v}_{at}\in\mathbb{C}^{N_a\times 1}$ is the transmit beamforming vector of CM, and $\textbf{w}_{a}\in\mathbb{C}^{N_a\times 1}$ denotes the beamforming vector for transmitting AN, where $\textbf{v}^H_{at}\textbf{v}_{at}=1$, and $\textbf{w}^H_{a}\textbf{w}_{a}=1$. $x_1$ is the CM with $\mathbb{E}[\|x_1\|^2]=1$.

The transmit signal from Bob is given by
\begin{align}\label{1}
\textbf{S}_b&=\sqrt{\beta_2P_b} \textbf{v}_{bt}x_2+\sqrt{(1-\beta_2)P_b}\textbf{w}_{b},
\end{align}
where $P_b$ represents the total transmit power, $\beta_2$ and $(1-\beta_2)$ are the PA factors of CM and AN, respectively. $\textbf{v}_{bt}\in\mathbb{C}^{N_b\times 1}$ is the transmit beamforming vector that sends CM, and $\textbf{w}_{b}\in\mathbb{C}^{N_b\times 1}$ denotes the AN beamforming vector, where $\textbf{v}^H_{bt}\textbf{v}_{bt}=1$, and $\textbf{w}^H_{b}\textbf{w}_{b}=1$. $x_2$ represents the CM with $\mathbb{E}[\|x_2\|^2]=1$.

Taking the path loss into account, the received signal at  Bob is
\begin{align}\label{y_b0}
y_{b}&=\textbf{v}^H_{br}\Big(\big(\sqrt{g_{ai_1b}}\textbf{H}^H_{i_1b}\bm{\Theta}_1\textbf{H}_{ai_1}+
\sqrt{g_{ai_2b}}\textbf{H}^H_{i_2b}\bm{\Theta}_2\textbf{H}_{ai_2}+\sqrt{g_{ab}}\textbf{H}^H_{ab})\textbf{S}_a+
\big(\sqrt{g_{bi_1b}}\textbf{H}^H_{i_1b}\bm{\Theta}_1\textbf{H}_{bi_1}+\nonumber\\
&~~\sqrt{g_{bi_2b}}\textbf{H}^H_{i_2b}\bm{\Theta}_2\textbf{H}_{bi_2}\big)\textbf{v}_{bt}x_2+\textbf{n}_{b}\Big)\nonumber\\
&=\textbf{v}^H_{br}\Big(\sqrt{\beta_1P_a} \textbf{H}_{b}(\bm{\Theta}_1, \bm{\Theta}_2)\textbf{v}_{at}x_1+
\big(\sqrt{g_{bi_1b}}\textbf{H}^H_{i_1b}\bm{\Theta}_1\textbf{H}_{bi_1}+
\sqrt{g_{bi_2b}}\textbf{H}^H_{i_2b}\bm{\Theta}_2\textbf{H}_{bi_2}\big)\textbf{v}_{bt}x_2+\nonumber\\
&~~\sqrt{(1-\beta_1)P_a}\textbf{H}_{b}(\bm{\Theta}_1, \bm{\Theta}_2)\textbf{w}_{a}+\textbf{n}_{b}\Big),
\end{align}
where
$\bm{\Theta}_1=\text{diag}(e^{j\phi^1_1}, \cdots, e^{j\phi^1_m}, \cdots, e^{j\phi^1_M})$ and $\bm{\Theta}_2=\text{diag}(e^{j\phi^2_1}, \cdots, e^{j\phi^2_m}, \cdots, e^{j\phi^2_M})$ are the diagonal reflection-coefficient matrices of RIS-1 and RIS-2, respectively, where $\phi^1_m, \phi^2_m \in [0,2\pi]$ represent the phase shifts of $m$-th reflection element. $\textbf{v}^H_{br}\in\mathbb{C}^{1\times N_{b}}$ is the receive beamforming vector.
$\textbf{n}_{b}\in \mathbb{C}^{N_{b}\times 1}$ is the complex additive white Gaussian noise (AWGN) vector with its distribution as $\textbf{n}_{b}\sim\mathcal {C}\mathcal {N}(0, \sigma^2_{b}\textbf{I}_{N_{b}})$. $g_{ai_1b}=g_{ai_1}g_{i_1b}$, $g_{ai_2b}=g_{ai_2}g_{i_2b}$, $g_{bi_1b}=g_{bi_1}g_{i_1b}$ and $g_{bi_2b}=g_{bi_2}g_{i_2b}$ represent the equivalent path loss coefficients of Alice-RIS-1-Bob, Alice-RIS-2-Bob, Bob-RIS-1-Bob and  Bob-RIS-2-Bob channels, respectively. $g_{ab}$ denotes the path loss coefficient of Alice-to-Bob channel. In (\ref{y_b0}),
\begin{align}\label{H_b}
\textbf{H}_{b}(\bm{\Theta}_1, \bm{\Theta}_2)
=&\sqrt{g_{ai_1b}}\textbf{H}^H_{i_1b}\bm{\Theta}_1\textbf{H}_{ai_1}+
\sqrt{g_{ai_2b}}\textbf{H}^H_{i_2b}\bm{\Theta}_2\textbf{H}_{ai_2}+\sqrt{g_{ab}}\textbf{H}^H_{ab},
\end{align}
where the channel matrices
$\textbf{H}_{ai_1}=\textbf{h}(\theta_{r,ai_1})\textbf{h}^H(\theta_{t,ai_1})\in\mathbb{C}^{M\times N_a}$,
$\textbf{H}^H_{i_1b}=\textbf{h}(\theta_{r,i_1b})\textbf{h}^H(\theta_{t,i_1b})\in\mathbb{C}^{N_{b}\times M}$,
$\textbf{H}_{ai_2}=\textbf{h}(\theta_{r,ai_2})\textbf{h}^H(\theta_{t,ai_2})\in\mathbb{C}^{M\times N_a}$, $\textbf{H}^H_{i_2b}=\textbf{h}(\theta_{r,i_2b})\textbf{h}^H(\theta_{t,i_2b})\in\mathbb{C}^{N_{b}\times M}$,
$\textbf{H}^H_{ab}=\textbf{h}(\theta_{r,ab})\textbf{h}^H(\theta_{t,ab})\in\mathbb{C}^{N_b\times N_a}$,
$\textbf{H}_{bi_1}=\textbf{h}(\theta_{r,bi_1})\textbf{h}^H(\theta_{t,bi_1})\in\mathbb{C}^{M\times N_b}$,
and $\textbf{H}_{bi_2}=\textbf{h}(\theta_{r,bi_2})\textbf{h}^H(\theta_{t,bi_2})\in\mathbb{C}^{M\times N_b}$
are the Alice-to-RIS-1, RIS-1-to-Bob, Alice-to-RIS-2, RIS-2-to-Bob, Alice-to-Bob, Bob-to-RIS-1, and Bob-to-RIS-2 channels, respectively.
The normalized steering vector $\textbf{h}(\theta)$ is
\begin{align}\label{h_theta}
\textbf{h}(\theta)=\frac{1}{\sqrt{N}}[e^{j2\pi\Psi_{\theta}(1)}, \dots, e^{j2\pi\Psi_{\theta}(n)}, \dots, e^{j2\pi\Psi_{\theta}(N)}]^T,
\end{align}
and the phase function $\Psi_{\theta}(n)$ is given by
\begin{align}
\Psi_{\theta}(n)\buildrel \Delta \over =-\frac{(n-(N+1)/2)d \cos\theta}{\lambda}, n=1, \dots, N,
\end{align}
where $\theta$ represents the direction angle of arrival or departure, $n$ denotes the index of antenna, $d$ is the  spacing of adjacent transmitting antennas, and $\lambda$ stands for the wavelength. Then (\ref{H_b}) can be rewritten as
\begin{align}\label{H_b1}
\textbf{H}_{b}(\bm{\Theta}_1,\bm{\Theta}_2)=&\sqrt{g_{ai_1b}}\textbf{h}(\theta_{r,i_1b})\textbf{h}^H(\theta_{t,i_1b})\bm{\Theta}_1
\textbf{h}(\theta_{r,ai_1})\textbf{h}^H(\theta_{t,ai_1})
+\sqrt{g_{ai_2b}}\textbf{h}(\theta_{r,i_2b})\textbf{h}^H(\theta_{t,i_2b})\bm{\Theta}_2\cdot\nonumber\\
&\textbf{h}(\theta_{r,ai_2})\textbf{h}^H(\theta_{t,ai_2})+\sqrt{g_{ab}}\textbf{h}(\theta_{r,ab})\textbf{h}^H(\theta_{t,ab}).
\end{align}
Assuming that the channel state information (CSI) of each link is perfectly known by Bob, similar to \cite{Shen2020Beamforming}, the term in (\ref{y_b0})
\begin{align*}  
\textbf{v}^H_{br}\left(\sqrt{g_{bi_1b}}\textbf{H}^H_{i_1b}\bm{\Theta}_1\textbf{H}_{bi_1}+
\sqrt{g_{bi_2b}}\textbf{H}^H_{i_2b}\bm{\Theta}_2\textbf{H}_{bi_2}\right)\textbf{v}_{bt}x_2
\end{align*}
can be removed from the received signal $y_b$ due to the fact that Bob knows its own data symbol $x_2$.
Then the received signal (\ref{y_b0}) reduces to
\begin{align}\label{y_b1}
y_{b}&=\textbf{v}^H_{br}\Big(\sqrt{\beta_1P_a} \textbf{H}_{b}(\bm{\Theta}_1, \bm{\Theta}_2)
\textbf{v}_{at}x_1+\underbrace{\sqrt{(1-\beta_1)P_a}\textbf{H}_{b}(\bm{\Theta}_1, \bm{\Theta}_2)\textbf{w}_{a}+\textbf{n}_{b}}_{\bar{\textbf{n}}_{b}}\Big).
\end{align}

Similar to the received signal at Bob, Alice knows its own data symbol $x_1$. Then the received signal at Alice is given by
\begin{align}\label{y_a0}
y_{a}&=\textbf{v}^H_{ar}\Big(\big(\sqrt{g_{ai_1b}}\textbf{H}^H_{i_1a}\bm{\Theta}_1\textbf{H}_{bi_1}+
\sqrt{g_{ai_2b}}\textbf{H}^H_{i_2a}\bm{\Theta}_2\textbf{H}_{bi_2}+\sqrt{g_{ab}}\textbf{H}^H_{ba}\big)\textbf{S}_b+\textbf{n}_{a}\Big)\nonumber\\
&=\textbf{v}^H_{ar}\Big(\sqrt{\beta_2P_b}\textbf{H}_{a}(\bm{\Theta}_1,\bm{\Theta}_2)\textbf{v}_{bt}x_2+
\underbrace{\sqrt{(1-\beta_2)P_b}\textbf{H}_{a}(\bm{\Theta}_1, \bm{\Theta}_2)
\textbf{w}_{b}+\textbf{n}_{a}}_{\bar{\textbf{n}}_{a}}\Big),
\end{align}
where
\begin{align}\label{H_a}
\textbf{H}_{a}(\bm{\Theta}_1, \bm{\Theta}_2)
=&\sqrt{g_{ai_1b}}\textbf{H}^H_{i_1a}\bm{\Theta}_1\textbf{H}_{bi_1}+
\sqrt{g_{ai_2b}}\textbf{H}^H_{i_2a}\bm{\Theta}_2\textbf{H}_{bi_2}+\sqrt{g_{ab}}\textbf{H}^H_{ba},
\end{align}
$\textbf{v}^H_{ar}\in\mathbb{C}^{1\times N_a}$ denotes the receive beamforming vector, $\textbf{n}_{a}\in \mathbb{C}^{N_{a}\times 1}$ is the complex AWGN vector,  distributed as $\textbf{n}_{a}\sim\mathcal {C}\mathcal {N}(0, \sigma^2_{a}\textbf{I}_{N_{a}})$,
the channel matrices
$\textbf{H}^H_{i_1a}=\textbf{h}(\theta_{r,i_1a})\textbf{h}^H(\theta_{t,i_1a})\in\mathbb{C}^{N_{a}\times M}$,
$\textbf{H}^H_{i_2a}=\textbf{h}(\theta_{r,i_2a})\textbf{h}^H(\theta_{t,i_2a})\in\mathbb{C}^{N_{a}\times M}$, and
$\textbf{H}^H_{ba}=\textbf{h}(\theta_{r,ba})\textbf{h}^H(\theta_{t,ba})\in\mathbb{C}^{N_a\times N_b}$ represent
the RIS-1-to-Alice, RIS-2-to-Alice, and Bob-to-Alice channels, respectively. Then (\ref{H_a}) becomes as
\begin{align}\label{H_a1}
\textbf{H}_{a}(\bm{\Theta}_1,\bm{\Theta}_2)=&\sqrt{g_{ai_1b}}\textbf{h}(\theta_{r,i_1a})\textbf{h}^H(\theta_{t,i_1a})\bm{\Theta}_1
\textbf{h}(\theta_{r,bi_1})\textbf{h}^H(\theta_{t,bi_1})
+\sqrt{g_{ai_2b}}\textbf{h}(\theta_{r,i_2a})\textbf{h}^H(\theta_{t,i_2a})\bm{\Theta}_2\cdot\nonumber\\
&\textbf{h}(\theta_{r,bi_2})\textbf{h}^H(\theta_{t,bi_2})+\sqrt{g_{ab}}\textbf{h}(\theta_{r,ba})\textbf{h}^H(\theta_{t,ba}).
\end{align}

The receive signal at Eve can be expressed as
\begin{align}\label{y_e1}
y_{e}&=\textbf{v}^H_{er}\Big(\big(\sqrt{g_{ai_1e}}\textbf{H}^H_{i_1e}\bm{\Theta}_1\textbf{H}_{ai_1}+
\sqrt{g_{ai_2e}}\textbf{H}^H_{i_2e}\bm{\Theta}_2\textbf{H}_{ai_2}+
\sqrt{g_{ae}}\textbf{H}^H_{ae}\big)\textbf{S}_a+
\big(\sqrt{g_{bi_1e}}\textbf{H}^H_{i_1e}\bm{\Theta}_1\textbf{H}_{bi_1}+\nonumber\\
&~~~\sqrt{g_{bi_2e}}\textbf{H}^H_{i_2e}\bm{\Theta}_2\textbf{H}_{bi_2}
+\sqrt{g_{be}}\textbf{H}^H_{be}\big)\textbf{S}_b +\textbf{n}_e\Big)\nonumber\\
&=\textbf{v}^H_{er}\Big(\sqrt{\beta_1P_a}\textbf{H}_{e_1}(\bm{\Theta}_1, \bm{\Theta}_2) \textbf{v}_{at}x_1+
\sqrt{\beta_2P_b}\textbf{H}_{e_2}(\bm{\Theta}_1,\bm{\Theta}_2)\textbf{v}_{bt}x_2+\sqrt{(1-\beta_1)P_a}\textbf{H}_{e_1}(\bm{\Theta}_1, \bm{\Theta}_2)\textbf{w}_{a}\nonumber\\
&~~~+\sqrt{(1-\beta_2)P_b}\textbf{H}_{e_2}(\bm{\Theta}_1,\bm{\Theta}_2)\textbf{w}_{b}+\textbf{n}_e\Big)\nonumber\\
&=\textbf{v}^H_{er}\Big(\sqrt{\beta_1P_a}\textbf{H}_{e_1}(\bm{\Theta}_1, \bm{\Theta}_2) \textbf{v}_{at}x_1+\sqrt{\beta_2P_b}\textbf{H}_{e_2}(\bm{\Theta}_1, \bm{\Theta}_2)
\textbf{v}_{bt}x_2+\bar{\textbf{n}}_e\Big),
\end{align}
where
\begin{align}\label{H_e1}
\textbf{H}_{e_1}(\bm{\Theta}_1, \bm{\Theta}_2)=&\sqrt{g_{ai_1e}}\textbf{H}^H_{i_1e}\bm{\Theta}_1\textbf{H}_{ai_1}+
\sqrt{g_{ai_2e}}\textbf{H}^H_{i_2e}\bm{\Theta}_2\textbf{H}_{ai_2}+\sqrt{g_{ae}}\textbf{H}^H_{ae},
\end{align}
\begin{align}\label{H_e2}
\textbf{H}_{e_2}(\bm{\Theta}_1, \bm{\Theta}_2)=&\sqrt{g_{bi_1e}}\textbf{H}^H_{i_1e}\bm{\Theta}_1\textbf{H}_{bi_1}+
\sqrt{g_{bi_2e}}\textbf{H}^H_{i_2e}\bm{\Theta}_2\textbf{H}_{bi_2}
+\sqrt{g_{be}}\textbf{H}^H_{be},
\end{align}
\begin{align}
\bar{\textbf{n}}_e&=\sqrt{(1-\beta_1)P_a}\textbf{H}_{e_1}(\bm{\Theta}_1, \bm{\Theta}_2)\textbf{w}_{a}+
 \sqrt{(1-\beta_2)P_b}
\textbf{H}_{e_2}(\bm{\Theta}_1,\bm{\Theta}_2)\textbf{w}_{b}+\textbf{n}_e,
\end{align}
$\textbf{v}^H_{er}\in\mathbb{C}^{1\times N_e}$ denotes the receive beamforming vector, $\textbf{n}_e\in \mathbb{C}^{N_e\times 1}$ represents the AWGN vector, distributed as $\textbf{n}_e\sim\mathcal {C}\mathcal {N}(0, \sigma^2_e\textbf{I}_{N_e})$. $g_{ai_1e}=g_{ai_1}g_{i_1e}$, $g_{ai_2e}=g_{ai_2}g_{i_2e}$, $g_{bi_1e}=g_{bi_1}g_{i_1e}$, and $g_{bi_2e}=g_{bi_2}g_{i_2e}$ denote the equivalent path loss coefficients of Alice-RIS-1-Eve, Alice-RIS-2-Eve, Bob-RIS-1-Eve, and Bob-RIS-2-Eve channels, respectively. $g_{ae}$ and $g_{be}$ are the path loss coefficients of Alice-to-Eve and Bob-to-Eve channels, respectively.  The channel matrices
$\textbf{H}^H_{ae}=\textbf{h}(\theta_{r,ae})\textbf{h}^H(\theta_{t,ae})\in\mathbb{C}^{N_e\times N_a}$,
$\textbf{H}^H_{be}=\textbf{h}(\theta_{r,be})\textbf{h}^H(\theta_{t,be})\in\mathbb{C}^{N_e\times N_b}$,
$\textbf{H}^H_{i_1e}=\textbf{h}(\theta_{r,i_1e})\textbf{h}^H(\theta_{t,i_1e})\in\mathbb{C}^{N_e\times M}$, and $\textbf{H}^H_{i_2e}=\textbf{h}(\theta_{r,i_2e})\textbf{h}^H(\theta_{t,i_2e})\in\mathbb{C}^{N_e\times M}$ represent the Alice-to-Eve, Bob-to-Eve, RIS-1-to-Eve, and RIS-2-to-Eve channels, respectively.  Then (\ref{H_e1}) and (\ref{H_e2}) can be rewritten as
\begin{align}
\textbf{H}_{e_1}(\bm{\Theta}_1, \bm{\Theta}_2)
=&\sqrt{g_{ai_1e}}\textbf{h}(\theta_{r,i_1e})\textbf{h}^H(\theta_{t,i_1e})\bm{\Theta}_1
\textbf{h}(\theta_{r,ai_1})\textbf{h}^H(\theta_{t,ai_1})
+\sqrt{g_{ai_2e}}\textbf{h}(\theta_{r,i_2e})\textbf{h}^H(\theta_{t,i_2e})\bm{\Theta}_2\cdot\nonumber\\
&\textbf{h}(\theta_{r,ai_2})\textbf{h}^H(\theta_{t,ai_2})+\sqrt{g_{ae}}\textbf{h}(\theta_{r,ae})\textbf{h}^H(\theta_{t,ae})
\end{align}
and
\begin{align}
\textbf{H}_{e_2}(\bm{\Theta}_1, \bm{\Theta}_2)
=&\sqrt{g_{bi_1e}}\textbf{h}(\theta_{r,i_1e})\textbf{h}^H(\theta_{t,i_1e})\bm{\Theta}_1
\textbf{h}(\theta_{r,bi_1})\textbf{h}^H(\theta_{t,bi_1})
+\sqrt{g_{bi_2e}}\textbf{h}(\theta_{r,i_2e})\textbf{h}^H(\theta_{t,i_2e})\bm{\Theta}_2\cdot\nonumber\\
&\textbf{h}(\theta_{r,bi_2})\textbf{h}^H(\theta_{t,bi_2})+\sqrt{g_{be}}\textbf{h}(\theta_{r,be})\textbf{h}^H(\theta_{t,be}),
\end{align}
respectively.

In this section, we characterize the SSR expression in this paper. According to formulas (\ref{y_a0}), (\ref{y_b1}), and (\ref{y_e1}), the achievable rates at Alice, Bob, and Eve are
\begin{align}\label{Ra}
&R_{a}=\text{log}_2\left(1+\frac{\textbf{v}^H_{ar}\textbf{A}\textbf{v}_{ar}}
{\textbf{v}^H_{ar}\textbf{B}\textbf{v}_{ar}+\sigma^2_{a}}\right),
\end{align}
\begin{align}\label{Rb}
&R_{b}=\text{log}_2\left(1+\frac{\textbf{v}^H_{br}\textbf{C}\textbf{v}_{br}}
{\textbf{v}^H_{br}\textbf{D}\textbf{v}_{br}+\sigma^2_{b}}\right),
\end{align}
and
\begin{align}\label{Re}
R_e=&\text{log}_2
\left(1+\frac{\textbf{v}^H_{er}\textbf{E}\textbf{v}_{er}}
{\textbf{v}^H_{er}(\textbf{G}+\textbf{J})\textbf{v}_{er}+\sigma^2_{e}}\right)+
\text{log}_2\left(1+\frac{\textbf{v}^H_{er}\textbf{F}\textbf{v}_{er}}
{\textbf{v}^H_{er}(\textbf{G}+\textbf{J})\textbf{v}_{er}+\sigma^2_{e}}\right),
\end{align}
respectively, where
\begin{align}
&\textbf{A}=\beta_2P_b\textbf{H}_{a}(\bm{\Theta}_1, \bm{\Theta}_2) \textbf{v}_{bt} \textbf{v}^H_{bt}\textbf{H}^H_{a}(\bm{\Theta}_1, \bm{\Theta}_2),
\textbf{B}=(1-\beta_2)P_b\textbf{H}_{a}(\bm{\Theta}_1, \bm{\Theta}_2)\textbf{w}_{b}
\textbf{w}^H_{b}\textbf{H}^H_{a}(\bm{\Theta}_1, \bm{\Theta}_2),\nonumber
\end{align}
\begin{align}
&\textbf{C}=\beta_1P_a\textbf{H}_{b}(\bm{\Theta}_1, \bm{\Theta}_2) \textbf{v}_{at} \textbf{v}^H_{at}\textbf{H}^H_{b}(\bm{\Theta}_1, \bm{\Theta}_2),
\textbf{D}=(1-\beta_1)P_a\textbf{H}_{b}(\bm{\Theta}_1, \bm{\Theta}_2)\textbf{w}_{a}
\textbf{w}^H_{a}\textbf{H}^H_{b}(\bm{\Theta}_1, \bm{\Theta}_2),\nonumber\\
&\textbf{E}= \beta_1P_a\textbf{H}_{e_1}(\bm{\Theta}_1, \bm{\Theta}_2) \textbf{v}_{at}
\textbf{v}^H_{at}\textbf{H}^H_{e_1}(\bm{\Theta}_1, \bm{\Theta}_2) ,
\textbf{F}= \beta_2P_b \textbf{H}_{e_2}(\bm{\Theta}_1,\bm{\Theta}_2)\textbf{v}_{bt}
\textbf{v}^H_{bt}\textbf{H}^H_{e_2}(\bm{\Theta}_1, \bm{\Theta}_2) ,\nonumber\\
&\textbf{G}=\left(1-\beta_1\right)P_a\textbf{H}_{e_1}(\bm{\Theta}_1, \bm{\Theta}_2)\textbf{w}_{a}
\textbf{w}^H_{a}\textbf{H}^H_{e_1}(\bm{\Theta}_1, \bm{\Theta}_2),
\textbf{J}=\left(1-\beta_2\right)P_b \textbf{H}_{e_2}(\bm{\Theta}_1, \bm{\Theta}_2)\textbf{w}_{b}
\textbf{w}^H_{b}\textbf{H}^H_{e_2}(\bm{\Theta}_1, \bm{\Theta}_2).
\end{align}
Then the achievable SSR can be written as
\begin{align}\label{p0}
R=\text{max}\{ 0, R_{a}+R_{b}- R_e\}.
\end{align}

\section{Proposed transmit beamforming methods}
In this section, the GPG method is proposed to design the RIS phase-shifting firstly. Then two transmit beamforming methods at Alice and Bob, called Max-SV and Max-SLNR, are presented to enhance the SSR performance by fully exploiting the double-RIS.
\subsection{Proposed GPG method of synthesizing  the phase-shifting matrices at two RISs}
%

Observing (\ref{H_b1}) and (\ref{H_a1}), it is obvious that their first terms on the right sides are the linear functions of  the RIS-1 phase-shifting matrix. To make a good balance between Alice and Bob, it is fairly reasonable to maximize the power sum of the two terms by a detailed design of $\bm{\Theta}_1$
\begin{align}
&\text{tr}(\textbf{h}^H(\theta_{t,i_1b})\bm{\Theta}_1\textbf{h}(\theta_{r,ai_1}))+
\text{tr}(\textbf{h}^H(\theta_{t,i_1a})\bm{\Theta}_1\textbf{h}(\theta_{r,bi_1}))\nonumber\\
&=\text{tr}(\bm{\Theta}_1\textbf{h}(\theta_{r,ai_1})\textbf{h}^H(\theta_{t,i_1b}))+
\text{tr}(\bm{\Theta}_1\textbf{h}(\theta_{r,bi_1})\textbf{h}^H(\theta_{t,i_1a}))\nonumber\\
&=\text{tr}(\bm{\Theta}_1(\textbf{h}(\theta_{r,ai_1})\textbf{h}^H(\theta_{t,i_1b})+
\textbf{h}(\theta_{r,bi_1})\textbf{h}^H(\theta_{t,i_1a})))\nonumber\\
&=\frac{1}{M}\sum\limits_{m = 1}^M
e^{j\phi^1_m}\Big(e^{j2\pi\left(\Psi_{\theta_{r,ai_1}}(m)-\Psi_{\theta_{t,i_1b}}(m)\right)}+
e^{j2\pi\left(\Psi_{\theta_{r,bi_1}}(m)-\Psi_{\theta_{t,i_1a}}(m)\right)}\Big).
\end{align}
\begin{figure}[htbp] 
\centering
\includegraphics[width=0.4\textwidth]{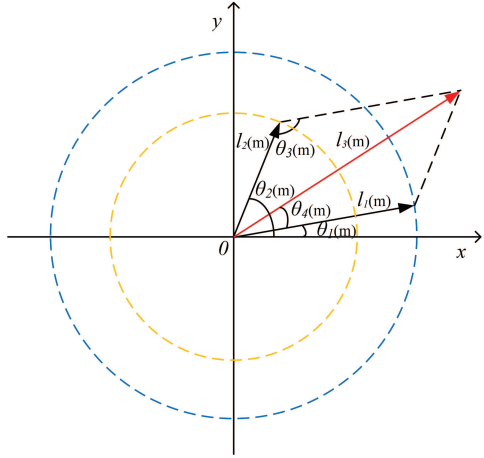}\\
\caption{Diagram of RIS-1 phase-shifting matrix designed.}\label{cycle.eps}
\end{figure}
To make more clear, as shown in Fig.~\ref{cycle.eps}, let us define
\begin{align}
&\theta_1(m)=2\pi(\Psi_{\theta_{r,ai_1}}(m)-\Psi_{\theta_{t,i_1b}}(m)),\\
&\theta_2(m)=2\pi(\Psi_{\theta_{r,bi_1}}(m)-\Psi_{\theta_{t,i_1a}}(m)).
\end{align}
In accordance with the angle relationship of parallelogram, we have
\begin{align}
&\theta_3(m)=\pi-\theta_2(m)+\theta_1(m).
\end{align}
Based on the cosine theorem, it is known that
\begin{align}
&l_3(m)=\sqrt{l^2_1(m)+l^2_2(m)-2l_1(m)l_2(m)\cos(\theta_3(m))} \label{L3},\\
&\theta_4(m)=\arccos \left(\frac{l^2_1(m)+l^2_3(m)-l^2_2(m)}{2l_1(m)l_3(m)}\right) \label{theta_4},
\end{align}
where $l_1(m)$ and $l_2(m)$ represent the weight coefficients of $\theta_1(m)$ and $\theta_2(m)$, respectively. Then we can obtain that
\begin{align}
l_3(m)e^{j(\theta_1(m)+\theta_4(m))}e^{j\phi^1_m}=c(m),
\end{align}
where $c(m)$ is a constant, $l_3(m)$ represents the weight coefficient of $\phi^1_m$. When $l_1(m)=l_2(m)$, (\ref{L3}) and (\ref{theta_4}) can be reduced to
\begin{align}
l_3(m)=l_1(m)\sqrt{2-2\cos(\theta_3(m))}
\end{align}
and
\begin{align}
\theta_4(m)=\frac{|\theta_2(m)-\theta_1(m)|}{2},
\end{align}
respectively. To maximize the power sum via RIS-1, let us set
\begin{align}
\theta_1(m)+\theta_4(m)+\phi^1_m=0,
\end{align}
then the optimal phase-shifting of $m$-th reflection element of $\bm{\Theta}_1$ is given by
\begin{align}
\phi^1_m=-(\theta_1(m)+\theta_4(m)).
\end{align}

Similarly, the optimal phase-shifting matrix of $\bm{\Theta}_2$ can also be obtained.

%
\subsection{Proposed Max-SV method}
In this section, a Max-SV beamforming method is proposed. Here, we first perform singular-value decomposition (SVD) on the desired channel from Alice to Bob. Its right singular vector corresponding to the largest singular-value  is used as the transmit beamforming vector. In the same manner, the receive beamforming vector is designed. The AN beamforming vector is constructed to maximizing the receive power at Eve on the null-space of this singular vector.

\subsubsection{Design of the CM beamforming vectors at Alice and Bob}

According to (\ref{y_b1}), the receive signal at Bob can be rewritten as
\begin{align}
y_{b}&=\textbf{v}^H_{br}\left(\sqrt{\beta_1P_a} \textbf{H}_{b}(\bm{\Theta}_1, \bm{\Theta}_2)
\textbf{v}_{at}x_1+\bar{\textbf{n}}_{b}\right).
\end{align}
Here, $\textbf{H}_{b}(\bm{\Theta}_1, \bm{\Theta}_2)$  has the following  SVD form
\begin{align}
\textbf{H}_{b}(\bm{\Theta}_1, \bm{\Theta}_2)
=\textbf{U}_{\textbf{H}_{b}}\Sigma_{\textbf{H}_{b}}\textbf{V}^H_{\textbf{H}_{b}},
\end{align}
where both of $\textbf{U}_{\textbf{H}_{b}}\in \mathbb{C}^{N_b\times N_b}$ and $\textbf{V}_{\textbf{H}_{b}}\in \mathbb{C}^{N_a\times N_a}$ are unitary matrices, and $\Sigma_{\textbf{H}_{b}}\in \mathbb{C}^{N_b\times N_a}$ is a matrix containing the singular values of $\textbf{H}_{b}(\bm{\Theta}_1, \bm{\Theta}_2)$ and along its main diagonal. The transmit beamforming vector $\textbf{v}_{at}$ and receive beamforming vector $\textbf{v}_{br}$ can be chosen as
\begin{align}
\textbf{v}_{at}=\textbf{V}_{\textbf{H}_{b}}(:,1)
\end{align}
and
\begin{align}
\textbf{v}_{br}=\textbf{U}_{\textbf{H}_{b}}(:,1),
\end{align}
respectively, where $\textbf{V}_{\textbf{H}_{b}}(:,1)$ and $\textbf{U}_{\textbf{H}_{b}}(:,1)$ respectively denote the first column vectors of the matrices $\textbf{V}_{\textbf{H}_{b}}$ and $\textbf{U}_{\textbf{H}_{b}}$.

In the same manner, the SVD form of $\textbf{H}_{a}(\bm{\Theta}_1, \bm{\Theta}_2)$ in (\ref{y_a0}) is
\begin{align}
\textbf{H}_{a}(\bm{\Theta}_1, \bm{\Theta}_2)
=\textbf{U}_{\textbf{H}_{a}}\Sigma_{\textbf{H}_{a}}\textbf{V}^H_{\textbf{H}_{a}},
\end{align}
where $\textbf{U}_{\textbf{H}_{a}}\in \mathbb{C}^{N_a\times N_a}$ and $\textbf{V}_{\textbf{H}_{a}}\in \mathbb{C}^{N_b\times N_b}$ are unitary matrices, and $\Sigma_{\textbf{H}_{a}}\mathbb{C}^{N_a\times N_b}$ is a matrix containing the singular values of $\textbf{H}_{a}(\bm{\Theta}_1, \bm{\Theta}_2)$ and along its main diagonal. The transmit beamforming vector $\textbf{v}_{bt}$ and receive beamforming vector $\textbf{v}_{ar}$ can be respectively designed as
\begin{align}
\textbf{v}_{bt}=\textbf{V}_{\textbf{H}_{a}}(:,1),
\textbf{v}_{ar}=\textbf{U}_{\textbf{H}_{a}}(:,1).
\end{align}


\subsubsection{Design the AN transmit beamforming vectors}
To reduce the effect of AN on the desired users, we limit the AN into the null-space of CM transmit space, and then the AN beamforming vector $\textbf{w}_a$ at Alice can be casted as
\begin{align}\label{w_a}
\textbf{w}_a=\textbf{T}_{a}\textbf{u}_{a},
\end{align}
where $\textbf{T}_{a}\in \mathbb{C}^{N_a\times N_a}$, $\textbf{u}_{a}\in \mathbb{C}^{N_a\times 1}$ satisfies $\textbf{u}^H_{a}\textbf{u}_{a}=1$. In other words,
to minimize the AN power received by Bob, the $\textbf{T}_a$ is a projector on the null-space of CM transmit beamforming vector at Alice constructed as follows
\begin{align}\label{T_a}
\textbf{T}_{a}=\textbf{I}_{N_a}-\textbf{V}_{\textbf{H}_b}(:,1)\textbf{V}_{\textbf{H}_b}(:,1)^H.
\end{align}
The problem of  maximizing the AN power received by Eve is formulated as
\begin{align}
&\max \limits_{\textbf{u}_{a}}
~~~\text{tr}\{\textbf{w}^H_{a}\textbf{H}_{ae}\textbf{H}^H_{ae}\textbf{w}_{a}\}
~~~\text{s.t.}~~\textbf{u}^H_{a}\textbf{u}_{a}=1.
\end{align}
Considering $\textbf{H}^H_{ae}=\textbf{h}(\theta_{r,ae})\textbf{h}^H(\theta_{t,ae})$, the above optimization problem reduces to
\begin{align}
&\max \limits_{\textbf{u}_a}
~~~\text{tr}\{\textbf{u}^H_a\textbf{T}^H_a\textbf{h}(\theta_{t,ae})\textbf{h}^H(\theta_{t,ae})\textbf{T}_a\textbf{u}_a\}
~~~\text{s.t.}~~\textbf{u}^H_a\textbf{u}_a=1,
\end{align}
which gives the associated Lagrangian
\begin{align}
L(\textbf{u}_{a},\lambda_a)=\textbf{u}^H_{a}\textbf{T}^H_{a}\textbf{h}(\theta_{t,ae})\textbf{h}^H(\theta_{t,ae})
\textbf{T}_{a}\textbf{u}_{a}-\lambda_a(\textbf{u}^H_{a}\textbf{u}_{a}-1),
\end{align}
where $\lambda_a$ is the Lagrange multiplier. We have the partial derivative of the Lagrangian function with respect to $\textbf{u}^*_{a}$ and set as zero
\begin{align}
\frac{\partial L(\textbf{u}_{a},\lambda_a)}{\partial{\textbf{u}^*_{a}}}
=(\textbf{T}^H_{a}\textbf{h}(\theta_{t,ae})\textbf{h}^H(\theta_{t,ae})\textbf{T}_{a})\textbf{u}_{a}-
\lambda_a\textbf{u}_{a}=0,
\end{align}
which is rewritten as
\begin{align}
\underbrace{\textbf{T}^H_{a}\textbf{h}(\theta_{t,ae})}_{\bf{t}}
\underbrace{\textbf{h}^H(\theta_{t,ae})\textbf{T}_{a}\textbf{u}_{a}}_{c}=c\bf{t}=
\lambda_a\textbf{u}_{a},
\end{align}
which means that $\textbf{u}_{a}$ is on the subspace spanned by the column vector $\bf{t}$ directly given by
\begin{align}\label{u_a}
\textbf{u}_{a}=\frac{\textbf{T}^H_{a}\textbf{h}(\theta_{t,ae})}{\|\textbf{T}^H_{a}\textbf{h}(\theta_{t,ae})\|},
\end{align}
Plugging (\ref{T_a}) and (\ref{u_a}) into (\ref{w_a}), the AN beamforming vector $\textbf{w}_a$ can be obtained completely.

Similarly, the AN beamforming vector $\textbf{w}_{b}$ at Bob is given by
\begin{align}
\textbf{w}_b=\textbf{T}_{b}\textbf{u}_{b},
\end{align}
where
\begin{align}
\textbf{T}_{b}=\textbf{I}_{N_b}-\textbf{V}_{\textbf{H}_{b}}(:,1)\textbf{V}_{\textbf{H}_{b}}(:,1)^H, 
\textbf{u}_{b}=\frac{\textbf{T}^H_{b}\textbf{h}(\theta_{t,be})}{\|\textbf{T}^H_{b}\textbf{h}(\theta_{t,be})\|}.
\end{align}

\subsubsection{Proposed ZF-based MRC receive beamforming method at Eve}
Seeing Fig.~\ref{System-model}, Eve may eavesdrop four-way signals from RIS-1, RIS-2, Alice and Bob. The four-way signals interfere with each other. It is very necessary for Eve to separate them and then combine them coherently. Below, the ZF receive beamforming method is first presented to separate them, and the MRC is adopted to combine their separate versions.

To completely cancel the interference among four-way signals, the total receive beamforming vector is decomposed as
\begin{align}\label{ZF-MRC}
\textbf{v}^H_{er}\buildrel \Delta \over=\underbrace{\left[w_{e_1}\ w_{e_2} \ w_{e_3}  \ w_{e_4}\right]}_{\text{MRC}}\cdot
\underbrace{\left[ {\begin{array}{*{20}{c}}
\textbf{v}^H_{er_1}\\
\textbf{v}^H_{er_2}\\
\textbf{v}^H_{er_3}\\
\textbf{v}^H_{er_4}
\end{array}} \right]}_{\text{ZF}},
\end{align}
where $w_{e_{1}}, w_{e_{2}}$, $w_{e_3}$, and $w_{e_4}$ are the weight coefficients of MRC, $\textbf{v}^H_{er_{1}}$, $\textbf{v}^H_{er_{2}}$, $\textbf{v}^H_{er_{3}}$, $ \textbf{v}^H_{er_{4}} \in \mathbb{C}^{1\times N_e}$  are the receive sub-beamforming vectors of ZF. Substituting (\ref{ZF-MRC}) in  (\ref{y_e1}) yields
\begin{align}
y_{e}&=\textbf{v}^H_{er}\Big(\sqrt{\beta_1P_a}\textbf{H}_{e_1}(\bm{\Theta}_1, \bm{\Theta}_2) \textbf{v}_{at}x_1+\sqrt{\beta_2P_b}\textbf{H}_{e_2}(\bm{\Theta}_1, \bm{\Theta}_2)
\textbf{v}_{bt}x_2+\bar{\textbf{n}}_e\Big)\nonumber\\
&=\left(w_{e_1}\textbf{v}^H_{er_1}+w_{e_2}\textbf{v}^H_{er_2}+ w_{e_3}\textbf{v}^H_{er_3}+w_{e_4}\textbf{v}^H_{er_4}\right)\Big[\textbf{h}(\theta_{r,i_1e})
\textbf{h}^H(\theta_{t,i_1e})\bm{\Theta}_1\big(\sqrt{\beta_1P_ag_{ai_1e}}\textbf{H}_{ai_1}\textbf{v}_{at}x_1+\nonumber\\
&\sqrt{\beta_2P_bg_{bi_1e}}\textbf{H}_{bi_1}\textbf{v}_{bt}x_2\big)
+\textbf{h}(\theta_{r,i_2e})\textbf{h}^H(\theta_{t,i_2e})\bm{\Theta}_2
\big(\sqrt{\beta_1P_ag_{ai_2e}}\textbf{H}_{ai_2}\textbf{v}_{at}x_1+
\sqrt{\beta_2P_bg_{bi_2e}}\textbf{H}_{bi_2}\textbf{v}_{bt}x_2\big)\nonumber\\
&+\sqrt{\beta_1P_ag_{ae}}\textbf{H}^H_{ae} \textbf{v}_{at}x_1+\sqrt{\beta_2P_bg_{be}}
\textbf{H}^H_{be}
\textbf{v}_{bt}x_2\Big]+\left(w_{e_1}\textbf{v}^H_{er_1}+w_{e_2}\textbf{v}^H_{er_2} + w_{e_3}\textbf{v}^H_{er_3}+ w_{e_4}\textbf{v}^H_{er_4}\right)\bar{\textbf{n}}_e.
\end{align}
To design the sub-beamforming vector $\textbf{v}^H_{er_1}$, it  is assumed that $\textbf{h}(\theta_{r,i_1e})$ is the only one useful channel for $\textbf{v}^H_{er_1}$ to receive the reflected CM from RIS-1 and the remaining channels are useless, i.e., $\textbf{v}^H_{er_1}$ satisfies
\begin{align}
\textbf{v}^H_{er_1}\textbf{h}(\theta_{r,i_2e})=0,~ \textbf{v}^H_{er_1}\textbf{h}(\theta_{r,ae})=0, ~ \textbf{v}^H_{er_1}\textbf{h}(\theta_{r,be})=0,
\end{align}
the actual CM channel can be defined as
\begin{align}
&\textbf{H}_{er_{-1}}=\left[ \begin{array}{*{20}{c}}
\textbf{h}^H(\theta_{r,i_2e})\\
\textbf{h}^H(\theta_{r,ae})\\
\textbf{h}^H(\theta_{r,be})
\end{array}\right],
\end{align}
then $\textbf{v}_{er_1}$ can be set as
\begin{align}
\textbf{v}_{er_1}=\left(\textbf{I}_{N_e}-\textbf{H}^H_{er_{-1}}[\textbf{H}_{er_{-1}}\textbf{H}^H_{er_{-1}}]^{\dagger}
\textbf{H}_{er_{-1}}\right)\textbf{h}(\theta_{r,i_1e}).
\end{align}
Likewise, $\textbf{v}_{er_2}$, $\textbf{v}_{er_3}$, and $\textbf{v}_{er_4}$ are respectively set as follows
\begin{subequations}
\begin{align}
&\textbf{v}_{er_2}=\left(\textbf{I}_{N_e}-\textbf{H}^H_{er_{-2}}[\textbf{H}_{er_{-2}}\textbf{H}^H_{er_{-2}}]^{\dagger}
\textbf{H}_{er_{-2}}\right)\textbf{h}(\theta_{r,i_2e}),\\
&\textbf{v}_{er_3}=\left(\textbf{I}_{N_e}-\textbf{H}^H_{er_{-3}}[\textbf{H}_{er_{-3}}\textbf{H}^H_{er_{-3}}]^{\dagger}
\textbf{H}_{er_{-3}}\right)\textbf{h}(\theta_{r,ae}),\\
&\textbf{v}_{er_4}=\left(\textbf{I}_{N_e}-\textbf{H}^H_{er_{-4}}[\textbf{H}_{er_{-4}}\textbf{H}^H_{er_{-4}}]^{\dagger}
\textbf{H}_{er_{-4}}\right)\textbf{h}(\theta_{r,be}),
\end{align}
\end{subequations}
where
\begin{align}
&\textbf{H}_{er_{-2}}=\left[ \begin{array}{*{20}{c}}
\textbf{h}^H(\theta_{r,i_1e})\\
\textbf{h}^H(\theta_{r,ae})\\
\textbf{h}^H(\theta_{r,be})
\end{array}\right],
\textbf{H}_{er_{-3}}=\left[ \begin{array}{*{20}{c}}
\textbf{h}^H(\theta_{r,i_1e})\\
\textbf{h}^H(\theta_{r,i_2e})\\
\textbf{h}^H(\theta_{r,be})
\end{array}\right],
\textbf{H}_{er_{-4}}=\left[ \begin{array}{*{20}{c}}
\textbf{h}^H(\theta_{r,i_1e})\\
\textbf{h}^H(\theta_{r,i_2e})\\
\textbf{h}^H(\theta_{r,ae})
\end{array}\right].
\end{align}
According to the MRC rule, the weight coefficients $w_{e_1}, w_{e_2}, w_{e_3}, w_{e_4}$ are respectively given by
\begin{subequations}
\begin{align}
&w_{e_1}=\frac{\left(\textbf{v}^H_{er_1}\textbf{H}^H_{i_1e}\bm{\Theta}_1
\left(\sqrt{\beta_1P_ag_{ai_1e}}\textbf{H}_{ai_1}\textbf{v}_{at}+
\sqrt{\beta_2P_bg_{bi_1e}}\textbf{H}_{bi_1}\textbf{v}_{bt}\right)\right)^H}
{\|\textbf{v}^H_{er_1}\textbf{H}^H_{i_1e}\bm{\Theta}_1
\left(\sqrt{\beta_1P_ag_{ai_1e}}\textbf{H}_{ai_1}\textbf{v}_{at}+
\sqrt{\beta_2P_bg_{bi_1e}}\textbf{H}_{bi_1}\textbf{v}_{bt}\right)\|},\\
&w_{e_2}=\frac{\left(\textbf{v}^H_{er_2}\textbf{H}^H_{i_2e}\bm{\Theta}_2
\left(\sqrt{\beta_1P_ag_{ai_2e}}\textbf{H}_{ai_2}\textbf{v}_{at}+
\sqrt{\beta_2P_bg_{bi_2e}}\textbf{H}_{bi_2}\textbf{v}_{bt}\right)\right)^H}
{\|\textbf{v}^H_{er_2}\textbf{H}^H_{i_2e}\bm{\Theta}_2\left(\sqrt{\beta_1P_ag_{ai_2e}}\textbf{H}_{ai_2}\textbf{v}_{at}+
\sqrt{\beta_2P_bg_{bi_2e}}\textbf{H}_{bi_2}\textbf{v}_{bt}\right)\|},\\
&w_{e_3}=\frac{\left(\textbf{v}^H_{er_3}\textbf{H}^H_{ae}\textbf{v}_{at}\right)^H}
{\|\textbf{v}^H_{er_3}\textbf{H}^H_{ae}\textbf{v}_{at}\|},\\
&w_{e_4}=\frac{\left(\textbf{v}^H_{er_4}\textbf{H}^H_{be}\textbf{v}_{bt}\right)^H}
{\|\textbf{v}^H_{er_4}\textbf{H}^H_{be}\textbf{v}_{bt}\|}.
\end{align}
\end{subequations}
Then Eq. (\ref{y_e1}) can be further converted to
\begin{align}
y_{e}
&=w_{e_1}\textbf{v}^H_{er_1}\textbf{H}^H_{i_1e}\bm{\Theta}_1
\big(\sqrt{\beta_1P_ag_{ai_1e}}\textbf{H}_{ai_1}\textbf{v}_{at}x_1+\sqrt{\beta_2P_bg_{bi_1e}}\textbf{H}_{bi_1}\textbf{v}_{bt}x_2\big)
+w_{e_2}\textbf{v}^H_{er_2}\textbf{H}^H_{i_2e}\bm{\Theta}_2\cdot\nonumber\\
&~~\big(\sqrt{\beta_1P_ag_{ai_2e}}\textbf{H}_{ai_2}\textbf{v}_{at}x_1+
\sqrt{\beta_2P_bg_{bi_2e}}\textbf{H}_{bi_2}\textbf{v}_{bt}x_2\big)+
\sqrt{\beta_1P_ag_{ae}}w_{e_3}\textbf{v}^H_{er_3}\textbf{H}^H_{ae} \textbf{v}_{at}x_1+\nonumber\\
&~~\sqrt{\beta_2P_bg_{be}}w_{e_4}\textbf{v}^H_{er_4}\textbf{H}^H_{be}
\textbf{v}_{bt}x_2+\big(w_{e_1}\textbf{v}^H_{er_1}+w_{e_2}\textbf{v}^H_{er_2} +
w_{e_3}\textbf{v}^H_{er_3}+ w_{e_4}\textbf{v}^H_{er_4}\big)\bar{\textbf{n}}_e.
\end{align}
\subsection{Generalized leakage method}
In this section, the leakage concept in \cite{Tarighat2005A, Sadek2007A} is generalized to   design  the CM transmit beamforming vector and AN beamforming vector, and  called a generalized leakage (GL) in what follows.

\subsubsection{Design the CM transmit beamforming vector}

The $\textbf{H}_{ai_1}$, $\textbf{H}_{ai_2}$, and $\textbf{H}^H_{ab}$ channels can be viewed as the desired channels, while $\textbf{H}^H_{ae}$ viewed as the undesired channel.
In accordance with \cite{Tarighat2005A, Sadek2007A}, the transmit beamforming vector $\textbf{v}_{at}$ is designed by the optimization problem
\begin{subequations}\label{v_at}
\begin{align}\label{max-vri}
&\max \limits_{\textbf{v}_{at}}
~~\text{SLNR}(\textbf{v}_{at})\\
&~\text{s.t.}~~\textbf{v}^H_{at}\textbf{v}_{at}=1,
\end{align}
\end{subequations}
where
\begin{align}
&\text{SLNR}(\textbf{v}_{at})=\frac{\beta_1P_a\text{tr}\left\{\textbf{v}^H_{at}\left(g_{ai_1}\textbf{H}^H_{ai_1}
\textbf{H}_{ai_1}+g_{ai_2}\textbf{H}^H_{ai_2}\textbf{H}_{ai_2}+g_{ab}\textbf{H}_{ab}\textbf{H}^H_{ab}\right)
\textbf{v}_{at}\right\}}{\text{tr}\left\{\textbf{v}^H_{at}\left(\beta_1P_ag_{ae}\textbf{H}_{ae}\textbf{H}^H_{ae}+
\sigma^2_e\textbf{I}_{N_a}\right)\textbf{v}_{at}\right\}}.
\end{align}
According to the generalized Rayleigh-Ritz theorem \cite{Horn1987}, the transmit beamforming vector $\textbf{v}_{at}$ at Alice is directly equal to the eigen-vector corresponding to the largest eigenvalue of the matrix
\begin{align}\label{v_at1}
&\left[g_{ae}\textbf{H}_{ae}\textbf{H}^H_{ae}+(\beta_1P_a)^{-1}{\sigma^2_e}\textbf{I}_{N_a}\right]^{-1}
\left(g_{ai_1}\textbf{H}^H_{ai_1}\textbf{H}_{ai_1}+g_{ai_2}\textbf{H}^H_{ai_2}\textbf{H}_{ai_2}+
g_{ab}\textbf{H}_{ab}\textbf{H}^H_{ab}\right).
\end{align}

Similarly, the transmit beamforming vector $\textbf{v}_{bt}$ at Bob can be designed from the eigen-vector corresponding to the largest eigenvalue of the matrix
\begin{align}
&\left[g_{be}\textbf{H}_{be}\textbf{H}^H_{be}+(\beta_2P_b)^{-1}{\sigma^2_e}\textbf{I}_{N_b}\right]^{-1}
\left(g_{i_1b}\textbf{H}^H_{bi_1}\textbf{H}_{bi_1}+g_{i_2b}\textbf{H}^H_{bi_2}\textbf{H}_{bi_2}+
g_{ab}\textbf{H}_{ba}\textbf{H}^H_{ba}\right).
\end{align}

\subsubsection{Design the AN beamforming vector}
The $\textbf{H}^H_{ae}$ can be viewed as the desired channels, while $\textbf{H}_{ai_1}$, $\textbf{H}_{ai_2}$, and $\textbf{H}^H_{ab}$ channels are viewed as the undesired channel.
In the following, we compute the AN beamforming vector at Alice by the following maximizing leakage-AN-to-signal ratio (LANSR) optimization problem
\begin{align}
&\max \limits_{\textbf{w}_{a}}
~~\text{LANSR}(\textbf{w}_{a})
~~~\text{s.t.}~~\textbf{w}^H_{a}\textbf{w}_a=1,
\end{align}
where $\text{LANSR}(\textbf{w}_{a})$ is given by
\begin{align}\label{LANSR}
&\text{LANSR}(\textbf{w}_{a})=
\frac{(1-\beta_1)P_a\text{tr}\left\{g_{ae}\textbf{w}^H_{a}\textbf{H}_{ae}\textbf{H}^H_{ae}\textbf{w}_{a}\right\}}
{\text{tr}\left\{\textbf{w}^H_a
\left[(1-\beta_1)P_a\left(g_{ai_1}\textbf{H}^H_{ai_1}\textbf{H}_{ai_1}+
g_{ai_2}\textbf{H}^H_{ai_2}\textbf{H}_{ai_2}+g_{ab}\textbf{H}_{ab}\textbf{H}^H_{ab}\right)+\sigma^2_b\textbf{I}_{N_a}
\right]
\textbf{w}_a\right\}}.
\end{align}
Similar to (\ref{v_at})-(\ref{v_at1}), we have
\begin{align}
&\Big[\left(g_{ai_1}\textbf{H}^H_{ai_1}\textbf{H}_{ai_1}+
g_{ai_2}\textbf{H}^H_{ai_2}\textbf{H}_{ai_2}+g_{ab}\textbf{H}_{ab}\textbf{H}^H_{ab}\right)+
((1-\beta_1)P_a)^{-1}\sigma^2_b\textbf{I}_{N_a}\Big]^{-1}\cdot
\left(g_{ae}\textbf{H}_{ae}\textbf{H}^H_{ae}\right).
\end{align}

In the same manner, the AN beamforming vector $\textbf{w}_{b}$ at Bob is given by the eigen-vector corresponding to the largest eigenvalue of the matrix
\begin{align}
&\Big[\left(g_{i_1b}\textbf{H}^H_{bi_1}\textbf{H}_{bi_1}+
g_{i_2b}\textbf{H}^H_{bi_2}\textbf{H}_{bi_2}+g_{ab}\textbf{H}_{ba}\textbf{H}^H_{ba}\right)+
((1-\beta_2)P_b)^{-1}\sigma^2_a\textbf{I}_{N_b}\Big]^{-1}\cdot
\left(g_{be}\textbf{H}_{be}\textbf{H}^H_{be}\right).
\end{align}

\subsubsection{Design of the receive beamforming vector}
Considering that Bob receives three-way signals from RIS-1, RIS-2, and Alice, to combine them coherently,
similar to the design of receive beamforming at Eve in (\ref{ZF-MRC}), the ZF-based MRC receive beamforming method is still adopted as follows
\begin{align}\label{3-w-ZF-MRC}
\textbf{v}^H_{br}=\left[w_{b_1}\ w_{b_2} \ w_{b_3}\right]\cdot
\left[\textbf{v}^*_{br_1}\ \textbf{v}^*_{br_2} \ \textbf{v}^*_{br_3}\right]^T.
\end{align}
where the receive sub-beamforming vectors $\textbf{v}_{br_1}$, $\textbf{v}_{br_2}$, $\textbf{v}_{br_3}\in \mathbb{C}^{N_b\times 1}$ are  respectively given as follows
\begin{align}
&\textbf{v}_{br_1}=\left(\textbf{I}_{N_b}-\textbf{H}^H_{br_{-1}}[\textbf{H}_{br_{-1}}\textbf{H}^H_{br_{-1}}]^{\dagger}
\textbf{H}_{br_{-1}}\right)\textbf{h}(\theta_{r,i_1b}),
\textbf{v}_{br_2}=\left(\textbf{I}_{N_b}-\textbf{H}^H_{br_{-2}}[\textbf{H}_{br_{-2}}\textbf{H}^H_{br_{-2}}]^{\dagger}
\textbf{H}_{br_{-2}}\right)\textbf{h}(\theta_{r,i_2b}),\nonumber\\
&\textbf{v}_{br_3}=\left(\textbf{I}_{N_b}-\textbf{H}^H_{br_{-3}}[\textbf{H}_{br_{-3}}\textbf{H}^H_{br_{-3}}]^{\dagger}
\textbf{H}_{br_{-3}}\right)\textbf{h}(\theta_{r,ab}),
\end{align}
where
\begin{align}
\textbf{H}_{br_{-1}}=\left[ \begin{array}{*{20}{c}}
\textbf{h}^H(\theta_{r,i_2b})\\
\textbf{h}^H(\theta_{r,ab})
\end{array}\right],
\textbf{H}_{br_{-2}}=\left[ \begin{array}{*{20}{c}}
\textbf{h}^H(\theta_{r,i_1b})\\
\textbf{h}^H(\theta_{r,ab})
\end{array}\right],
\textbf{H}_{br_{-3}}=\left[ \begin{array}{*{20}{c}}
\textbf{h}^H(\theta_{r,i_1b})\\
\textbf{h}^H(\theta_{r,i_2b})
\end{array}\right].
\end{align}
In (\ref{3-w-ZF-MRC}),  the weight coefficients $w_{b_1}, w_{b_2}$, and $w_{b_3}$ can be respectively designed as follows
\begin{align}
&w_{b_1}=\frac{\left(\textbf{v}^H_{br_1}\textbf{H}^H_{i_1b}\bm{\Theta}_1\textbf{H}_{ai_1}\textbf{v}_{at}\right)^H}
{\|\textbf{v}^H_{br_1}\textbf{H}^H_{i_1b}\bm{\Theta}_1\textbf{H}_{ai_1}\textbf{v}_{at}\|},
w_{b_2}=\frac{\left(\textbf{v}^H_{br_2}\textbf{H}^H_{i_2b}\bm{\Theta}_2\textbf{H}_{ai_2}\textbf{v}_{at}\right)^H}
{\|\textbf{v}^H_{br_2}\textbf{H}^H_{i_2b}\bm{\Theta}_2\textbf{H}_{ai_2}\textbf{v}_{at}\|},
w_{b_3}=\frac{\left(\textbf{v}^H_{br_3}\textbf{H}^H_{ab}\textbf{v}_{at}\right)^H}
{\|\textbf{v}^H_{br_3}\textbf{H}^H_{ab}\textbf{v}_{at}\|}.
\end{align}

Therefore,  $(\ref{y_b1})$ can be further converted to
\begin{align}\label{y-b-w}
y_{b}=&\sqrt{\beta_1P_a}\Big(\sqrt{g_{ai_1b}}w_{b_1}\textbf{v}^H_{br_1} \textbf{h}(\theta_{r,i_1b})\textbf{h}^H(\theta_{t,i_1b})\bm{\Theta}_1\textbf{H}_{ai_1}+
\sqrt{g_{ai_2b}}w_{b_2}\textbf{v}^H_{br_2}\textbf{h}(\theta_{r,i_2b})\textbf{h}^H(\theta_{t,i_2b})
\bm{\Theta}_2\textbf{H}_{ai_2}+\nonumber\\
&\sqrt{g_{ab}}w_{b_3}\textbf{v}^H_{br_3}\textbf{h}(\theta_{r,ab})\textbf{h}^H(\theta_{t,ab})\Big)\textbf{v}_{at}x_1
+(w_{b_1}\textbf{v}^H_{br_1}+ w_{b_2}\textbf{v}^H_{br_2}+w_{b_3}\textbf{v}^H_{br_3})\bar{\textbf{n}}_{b}.
\end{align}

Similarly, the receive beamforming vector $\textbf{v}^H_{ar}$ at Alice is
\begin{align}\label{4-w-ZF-MRC}
\textbf{v}^H_{ar}=\left[w_{a_1}\ w_{a_2} \ w_{a_3}\right]\cdot
\left[\textbf{v}^*_{ar_1}\ \textbf{v}^*_{ar_2} \ \textbf{v}^*_{ar_3}\right]^T.
\end{align}
where the receive sub-beamforming vectors $\textbf{v}_{ar_1}$, $\textbf{v}_{ar_2}$, $\textbf{v}_{ar_3}\in \mathbb{C}^{N_a\times 1}$ are  respectively given by
\begin{align}
&\textbf{v}_{ar_1}=\left(\textbf{I}_{N_a}-\textbf{H}^H_{ar_{-1}}[\textbf{H}_{ar_{-1}}\textbf{H}^H_{ar_{-1}}]^{\dagger}
\textbf{H}_{ar_{-1}}\right)\textbf{h}(\theta_{r,i_1a}),\nonumber\\
&\textbf{v}_{ar_2}=\left(\textbf{I}_{N_a}-\textbf{H}^H_{ar_{-2}}[\textbf{H}_{ar_{-2}}\textbf{H}^H_{ar_{-2}}]^{\dagger}
\textbf{H}_{ar_{-2}}\right)\textbf{h}(\theta_{r,i_2a}),\nonumber\\
&\textbf{v}_{ar_3}=\left(\textbf{I}_{N_a}-\textbf{H}^H_{ar_{-3}}[\textbf{H}_{ar_{-3}}\textbf{H}^H_{ar_{-3}}]^{\dagger}
\textbf{H}_{ar_{-3}}\right)\textbf{h}(\theta_{r,ba}),
\end{align}
and
\begin{align}
\textbf{H}_{ar_{-1}}=\left[ \begin{array}{*{20}{c}}
\textbf{h}^H(\theta_{r,i_2a})\\
\textbf{h}^H(\theta_{r,ba})
\end{array}\right],
\textbf{H}_{ar_{-2}}=\left[ \begin{array}{*{20}{c}}
\textbf{h}^H(\theta_{r,i_1a})\\
\textbf{h}^H(\theta_{r,ba})
\end{array}\right],
\textbf{H}_{ar_{-3}}=\left[ \begin{array}{*{20}{c}}
\textbf{h}^H(\theta_{r,i_1a})\\
\textbf{h}^H(\theta_{r,i_2a})
\end{array}\right].
\end{align}
In (\ref{4-w-ZF-MRC}), the weight coefficients $w_{a_1}, w_{a_2}$, and $w_{a_3}$ are respectively constructed as follows
\begin{align}
&w_{a_1}=\frac{\left(\textbf{v}^H_{ar_1}\textbf{H}^H_{i_1a}\bm{\Theta}_1\textbf{H}_{bi_1}\textbf{v}_{bt}\right)^H}
{\|\textbf{v}^H_{ar_1}\textbf{H}^H_{i_1a}\bm{\Theta}_1\textbf{H}_{bi_1}\textbf{v}_{bt}\|},
w_{a_2}=\frac{\left(\textbf{v}^H_{ar_2}\textbf{H}^H_{i_2a}\bm{\Theta}_2\textbf{H}_{bi_2}\textbf{v}_{bt}\right)^H}
{\|\textbf{v}^H_{ar_2}\textbf{H}^H_{i_2a}\bm{\Theta}_2\textbf{H}_{bi_2}\textbf{v}_{bt}\|},
w_{a_3}=\frac{\left(\textbf{v}^H_{ar_3}\textbf{H}^H_{ba}\textbf{v}_{bt}\right)^H}
{\|\textbf{v}^H_{ar_3}\textbf{H}^H_{ba}\textbf{v}_{bt}\|}.
\end{align}
The received signal in (\ref{y_a0}) can be further converted to
\begin{align}
y_{a}&=\sqrt{\beta_2P_b}\Big(\sqrt{g_{ai_1b}}w_{a_1}\textbf{v}^H_{ar_1}\textbf{h}(\theta_{r,i_1a})
\textbf{h}^H(\theta_{t,i_1a})\bm{\Theta}_1\textbf{H}_{bi_1}+
\sqrt{g_{ai_2b}}w_{a_2}\textbf{v}^H_{ar_2}\textbf{h}(\theta_{r,i_2a})\textbf{h}^H(\theta_{t,i_2a})
\bm{\Theta}_2\textbf{H}_{bi_2}\nonumber\\
&~~+\sqrt{g_{ab}}w_{a_3}\textbf{v}^H_{ar_3}\textbf{h}(\theta_{r,ba})\textbf{h}^H(\theta_{t,ba})\Big)\textbf{v}_{bt}x_2+(w_{a_1}\textbf{v}^H_{ar_1} + w_{a_2}\textbf{v}^H_{ar_2} +w_{a_3}\textbf{v}^H_{ar_3})\bar{\textbf{n}}_{a}.
\end{align}
This completes the construction of all beamforming methods.

\section{Proposed HICF Power Allocation Strategy}

In this section, given that all beamforming vectors are designed well in the previous section,  we will optimize the PA between CM and AN to improve the SSR performance. The PA method of maximizing SSR is proposed. First, two exhaustive search (ES) methods including 2D and 1D are presented, and then a hybrid iterative and closed-form solution is proposed to reduce the high computational complexity and approximately achieve the same SSR performance as ES method.
\subsection{Problem formulation}
Given all beamforming vectors,    maximizing the SSR in (\ref{p0})  over the PA factors forms the following optimization problem
\begin{align}\label{R12}
&\max \limits_{\beta_1, \beta_2}
~~R(\beta_1,\beta_2)=R_{a}+R_{b}- R_e
~~~\text{s.t.} ~~0\leq\beta_1\leq1, 0\leq\beta_2\leq1.
\end{align}
Let us define
\begin{align}
&s_1=P_b\|\textbf{v}^H_{ar}\textbf{H}_{a}(\bm{\Theta}_1, \bm{\Theta}_2) \textbf{v}_{bt}\|^2,
s_2=P_b\|\textbf{v}^H_{ar}\textbf{H}_{a}(\bm{\Theta}_1, \bm{\Theta}_2)\textbf{w}_{b}\|^2,
s_3=P_a\|\textbf{v}^H_{br}\textbf{H}_{b}(\bm{\Theta}_1, \bm{\Theta}_2) \textbf{v}_{at}\|^2,\nonumber\\
&s_4=P_a\|\textbf{v}^H_{br}\textbf{H}_{b}(\bm{\Theta}_1, \bm{\Theta}_2)\textbf{w}_{a}\|^2,
s_5= P_a\|\textbf{v}^H_{er}\textbf{H}_{e_1}(\bm{\Theta}_1, \bm{\Theta}_2) \textbf{v}_{at}\|^2,
s_6= P_b \|\textbf{v}^H_{er}\textbf{H}_{e_2}(\bm{\Theta}_1,\bm{\Theta}_2)\textbf{v}_{bt}\|^2,\nonumber\\
&s_7=P_a\|\textbf{v}^H_{er}\textbf{H}_{e_1}(\bm{\Theta}_1, \bm{\Theta}_2)\textbf{w}_{a}\|^2,
s_8=P_b \|\textbf{v}^H_{er}\textbf{H}_{e_2}(\bm{\Theta}_1, \bm{\Theta}_2)\textbf{w}_{b}\|^2,
\end{align}
then the objective function $R(\beta_1,\beta_2)$ can be rewritten as follows
\begin{align}\label{R-1-2}
&R(\beta_1,\beta_2)
=\text{log}_2\left(1+\frac{\beta_2s_1}{(1-\beta_2)s_2+\sigma^2_a}\right)+
\text{log}_2\left(1+\frac{\beta_1s_3}{(1-\beta_1)s_4+\sigma^2_b}\right)-\nonumber\\
&\text{log}_2\left(1+\frac{\beta_1s_5}{(1-\beta_1)s_7+(1-\beta_2)s_8+\sigma^2_e}\right)-
\text{log}_2\left(1+\frac{\beta_2s_6}{(1-\beta_1)s_7+(1-\beta_2)s_8+\sigma^2_e}\right).
\end{align}
In what follows, let us consider two cases: $\beta_2\ne \beta_2$ (different, 2D) and $\beta_1=\beta_2$ (equal, 1D), which are called 2D-ES and 1D-ES, respectively.
\subsection{2D-ES and 1D-ES PA  strategies}
In this section, we first consider the case of $\beta_1\ne\beta_2$,  the 2D PA optimization problem in (\ref{R12}) can be recasted as
\begin{subequations}\label{R12-1}
\begin{align}
&\max \limits_{\beta_1, \beta_2}
~~R(\beta_1,\beta_2)=R_{a}+R_{b}- R_e\\
&~\text{s.t.} ~~~~~0\leq\beta_1\leq1, 0\leq\beta_2\leq1,
\beta_1\neq \beta_2.
\end{align}
\end{subequations}
Clearly, the above ojective function is a non-concave function. Due to its three constraints, it is hard to obtain its closed-form solution. It is  natural  to use a 2D-ES algorithm to find its approximate solution over the 2D domain $[0,1] \times [0,1]$.


To reduce the computational complexity of the above 2D-ES algorithm and consider the symmetry of two-way network,  $\beta_1$ is taken to be equal to $\beta_2$. Let us define $\beta_1=\beta_2=\beta$, then (\ref{R12}) reduces to
\begin{align}\label{R0-1D-ES}
&\max \limits_{\beta}
~~R(\beta)=\text{log}_2\frac{Q_1}{Q_2}
~~~\text{s.t.}~~~0\leq\beta_1, \beta_2\leq1,
\end{align}
where
\begin{align}
Q_1=&((s_1-s_2)\beta+s_2+\sigma^2_a)((s_3-s_4)\beta+s_4+\sigma^2_b)((-s_7-s_8)\beta+s_7+s_8+\sigma^2_e)^2,\\
Q_2=&(-s_2\beta+s_2+\sigma^2_a)(-s_4\beta+s_4+\sigma^2_b)((s_5-s_7-s_8)+s_7+ s_8+\sigma^2_e)\cdot\nonumber\\
&((s_6-s_7-s_8)\beta+s_7+ s_8+\sigma^2_e).
\end{align}
According to the derivation of Appendix A, (\ref{R0-1D-ES}) is equivalent to solving the following sixth-order polynomial
\begin{align}\label{bata^6}
f(\beta)=\beta^6+\alpha_1\beta^5+\alpha_2\beta^4+\alpha_3\beta^3+\alpha_4\beta^2+\alpha_5\beta+\alpha_6=0
\end{align}
with the constraint $\beta\in[0,1]$.

%
%
\subsection{Proposed HICF PA strategy}

\begin{figure}[htbp] 
\centering
\includegraphics[width=0.8\textwidth]{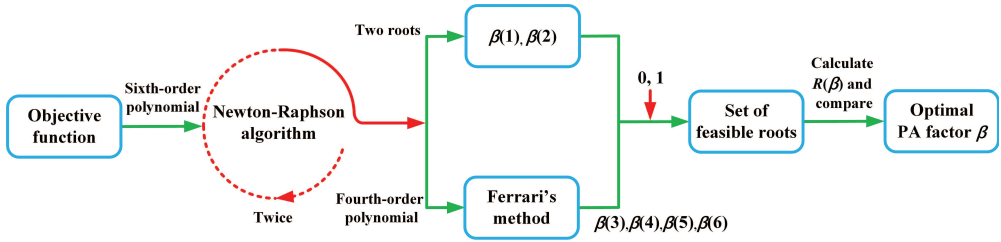}\\
\caption{Diagram for HICF power allocation strategy.}\label{step}
\end{figure}

To the best of our knowledge, there is no closed-form expression for roots of a general  polynomial with order more than four in (\ref{bata^6}). In what follows, we will propose  a HICF method to solve this polynomial, and its basic idea is  as follows: the Newton-Raphson algorithm in \cite{Wasserman2003All} is first employed twice to reduce its order  from six to four with two candidate roots be computed iteratively,  and the remaining four candidate roots can be obtained by the Ferrari's method.  The more detailed procedure are sketched in Fig.~\ref{step}.

Let us begin with the initialization of the Newton-Raphson method:
\begin{align}\label{f1}
f_1(\beta)=\beta^6+\alpha_1\beta^5+\alpha_2\beta^4+\alpha_3\beta^3+\alpha_4\beta^2+\alpha_5\beta+\alpha_6,
\end{align}
and its derivative
\begin{align}\label{g1}
g_1(\beta)=\frac{\partial f_1(\beta)}{\partial \beta}
=6\beta^5+5\alpha_1\beta^4+4\alpha_2\beta^3+3\alpha_3\beta^2+2\alpha_4\beta+\alpha_5.
\end{align}The iterative step of Newton-Raphson algorithm is as follows
\begin{align}\label{beta_p}
\beta^{p+1}=\beta^{p}-\frac{f_1(\beta^{p})}{g_1(\beta^{p})},
\end{align}
where $p$ is the number of iterations, and setting the initial value $\beta^{0}=0.5$. Repeating the above iterate process until $|\beta^{p+1}-\beta^{p}|\leq 10^{-5}$ yields the first  root  $\beta(1)$, and (\ref{bata^6})  is decomposed as a product of an one-order factor and one fifth-order factor as follows
\begin{align}
(\beta-\beta(1))
(\beta^5+\bar{\alpha}_1\beta^4+\bar{\alpha}_2\beta^3+\bar{\alpha}_3\beta^2+\bar{\alpha}_4\beta+\bar{\alpha}_5)=0,
\end{align}
where
\begin{align}
&\bar{\alpha}_1=\alpha_1+\beta(1), \bar{\alpha}_2=\alpha_2+\beta(1)\bar{\alpha}_1,
\bar{\alpha}_3=\alpha_3+\beta(1)\bar{\alpha}_2,\nonumber\\
&\bar{\alpha}_4=\alpha_4+\beta(1)\bar{\alpha}_3, \bar{\alpha}_5=\alpha_5+\beta(1)\bar{\alpha}_4.
\end{align}
The remaining five roots of (\ref{bata^6}) can be found by solving the roots of fifth-order polynomial
\begin{align}\label{beta^5}
\beta^5+\bar{\alpha}_1\beta^4+\bar{\alpha}_2\beta^3+\bar{\alpha}_3\beta^2+\bar{\alpha}_4\beta+\bar{\alpha}_5=0.
\end{align}
which is higher in order than four.  We still need to use the  Newton-Raphson algorithm one time.   Let us define the objective function and its derivative as
\begin{align}
f_2(\beta)=\beta^5+\bar{\alpha}_1\beta^4+\bar{\alpha}_2\beta^3+\bar{\alpha}_3\beta^2+\bar{\alpha}_4\beta+\bar{\alpha}_5,
\end{align}
and
\begin{align}
g_2(\beta)=\frac{\partial f_2(\beta)}{\partial \beta}=&
5\beta^4+4\bar{\alpha}_1\beta^3+3\bar{\alpha}_2\beta^2+2\bar{\alpha}_3\beta+\bar{\alpha}_4,
\end{align}
respectively. To avoid the increase of computational complexity caused by repeated search, we define a new reduced search domain initial value  $ (0, 0.5)\cup(\beta(1), 1)$ , and the initial value $\beta^{0}$ is randomly chosen in this interval. Repeating the procession of computing $\beta(1)$ in (\ref{beta_p}), an root $\beta(2)$ of (\ref{beta^5}) is obtained in the same manner, which is the second root of (\ref{bata^6}). Making use of the values of $\beta(1)$ and  $\beta(2)$, (\ref{bata^6}) has the following decomposition form
\begin{align}
&(\beta-\beta(1))(\beta-\beta(2))(\beta^4+\hat{\alpha}_1\beta^3+\hat{\alpha}_2\beta^2+\hat{\alpha}_3\beta+\hat{\alpha}_4)=0,
\end{align}
where
\begin{align}
\hat{\alpha}_1=\bar{\alpha}_1+\beta(2), \hat{\alpha}_2=\bar{\alpha}_2+\beta(2)\hat{\alpha}_1,
\hat{\alpha}_3=\bar{\alpha}_3+\beta(2)\hat{\alpha}_2, \hat{\alpha}_4=\bar{\alpha}_4+\beta(2)\hat{\alpha}_3.
\end{align}

Now, the two roots of (\ref{bata^6}) have been found. The problem of finding the remaining solutions can be converted to the one of solving the roots of the fourth-order polynomial as follows
\begin{align}\label{beta^4}
\beta^4+\hat{\alpha}_1\beta^3+\hat{\alpha}_2\beta^2+\hat{\alpha}_3\beta+\hat{\alpha}_4=0.
\end{align}
According to the Ferrari's method \cite{Cardano2007The}, the roots of (\ref{beta^4}) is given by
\begin{align}\label{beta_1}
\beta(3:6)=-\frac{\hat{\alpha}_1}{4}\pm_s\frac{\eta_1}{2}\pm_i\frac{\eta_2}{2},
\end{align}
where two $\pm_s$ have the same sign, while the sign of $\pm_i$ is independent,
\begin{align}
&\gamma_1=\frac{1}{3}(3\hat{\alpha}_1\hat{\alpha}_3-12\hat{\alpha}_4-\hat{\alpha}_2^2),
\gamma_2=\frac{1}{27}(-2\hat{\alpha}_2^3+9\hat{\alpha}_1\hat{\alpha}_2\hat{\alpha}_3+72\hat{\alpha}_2\hat{\alpha}_4
-27\hat{\alpha}_3^2-27\hat{\alpha}_1^2\hat{\alpha}_4),\nonumber\\
&\gamma_3=\frac{\hat{\alpha}_2}{3}+\sqrt[3]{-\frac{\gamma_2}{2}+\sqrt{\frac{\gamma_2^2}{4}+\frac{\gamma_1^3}{27}}}+
\sqrt[3]{-\frac{\gamma_2}{2}-\sqrt{\frac{\gamma_2^2}{4}+\frac{\gamma_1^3}{27}}},
\eta_1=\sqrt{\frac{\hat{\alpha}_1^2}{4}-\hat{\alpha}_2+\gamma_3},\nonumber\\
&\eta_2=\sqrt{\frac{3}{4}\hat{\alpha}_1^2-\eta_1^2-2\hat{\alpha}_2\pm_s\frac{1}{4\eta_1}
(4\hat{\alpha}_1\hat{\alpha}_2-8\hat{\alpha}_3-\hat{\alpha}_1^3)}.
\end{align}
At this point, all roots of the sixth-order polynomial in (\ref{bata^6}) have been found completely.  Then we have the set of all candidates for the optimal PA factor as
\begin{align}
S_{PA}=\{\beta(1), \beta(2), \beta(3), \beta(4), \beta(5), \beta(6), 0, 1\}.
\end{align}
The set of optimal values of $\beta$ is chosen from set $S_{PA}$ with two constraints: (1) falling in the interval $[0,1]$; (2) maximizing the SSR.

\section{Simulation Results and Discussions}
In this section, we make an evaluation on the performance of the proposed two transmit beamforming methods and one PA algorithm.
System parameters are given as follows:
$P_a=P_b=27$dBm, $N_a=N_b=N_e=8$, $M=100$, $d=\lambda/2$, $\beta_1=\beta_2=0.9$, $d_{ai_1}=d_{ai_2}=30$m, $d_{ab}=d_{ae}=80$m, $\theta_{t,ai_1}=\pi/8$, $\theta_{t,ai_2}=7\pi/8$, $\theta_{t,ae}=4\pi/9$, $\theta_{t,ab}=5\pi/9$,
$\sigma^2_a=\sigma^2_b=2\sigma^2_e$.
The path loss coefficient is defined as $g_{tr}=\frac{\alpha}{d^c_{tr}}$, where $\alpha$ is the path loss at reference distance $d_0$, $d_{tr}$ denotes the distance between the transmitter and receiver, and $c$ is the path loss exponent.

In what follows, three  schemes will be used as performance benchmarks:
\begin{enumerate}
\item Case I: \textbf{No RIS}: $\bm{\Theta}_1$=$\bm{\Theta}_2$=$\textbf{0}_{M\times M}$.
\item Case II: \textbf{RIS with random phase}: Phase of each element of  both $\bm{\Theta}_1$ and $\bm{\Theta}_2$ is uniformly and independently generated from the interval [0,2$\pi$).
\item Case III: \textbf{RIS-1/RIS-2}: let us set the phase-shifting matrix of one and only one of RIS-2 and RIS-1  as zero matrix.
\end{enumerate}

\begin{figure}[htbp]
\centering
\begin{minipage}[t]{0.45\textwidth}
\centering
\includegraphics[width=0.95\textwidth]{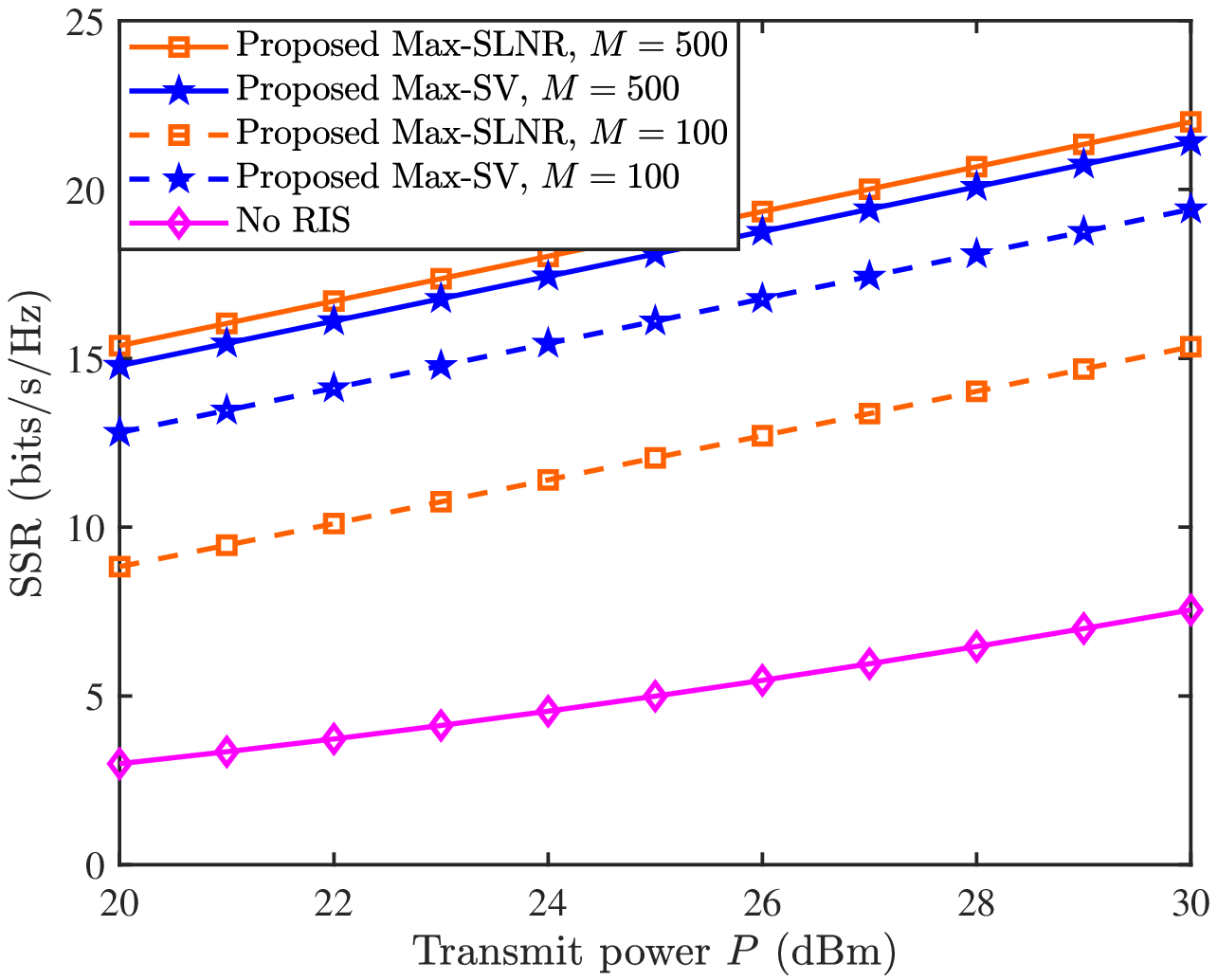}\\
\caption{Curves of SSR versus transmit power $P$ for different $M$.}\label{SSR-power}
\end{minipage}
\begin{minipage}[t]{0.45\textwidth}
\centering
\includegraphics[width=0.95\textwidth]{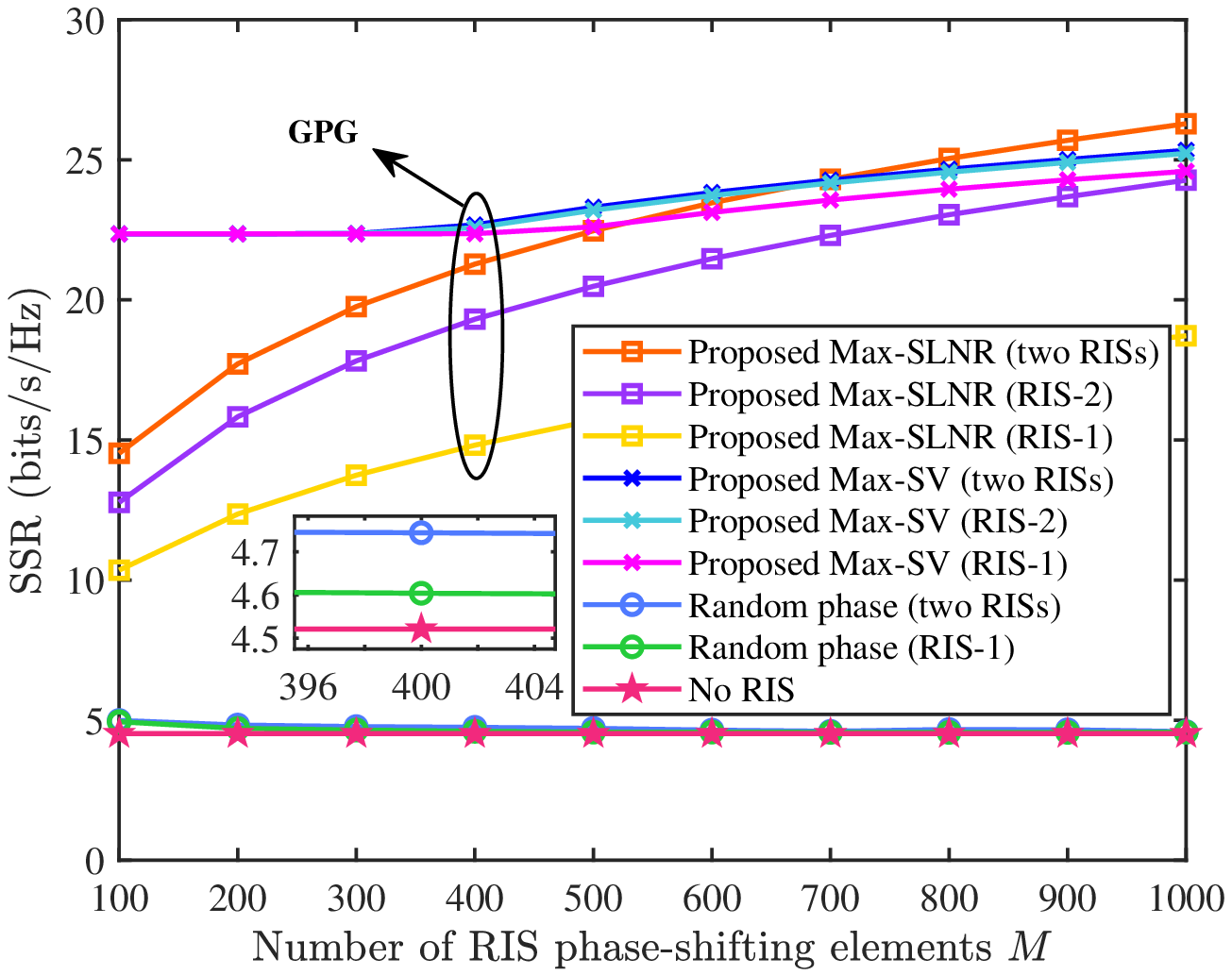}\\
\caption{Curves of SSR versus the number of RIS phase-shifting elements $M$ ($d_{ab}=70$ m).}\label{SSR-70}
\end{minipage}
\end{figure}

Fig.~{\ref{SSR-power}} demonstrates the curves of SSR versus transmit power $P$ with $P=P_a=P_b$ and  no RIS as a SSR performance benchmark. It can be seen from this figure that the proposed two methods Max-SV and Max-SLNR double and triple the SSR of no RIS at $M=100$ and  $M=500$, respectively. This means that double-RIS can bring a significant SSR improvement.

Fig.~{\ref{SSR-70}} plots the curves of SSR versus  the number $M$ of RIS phase-shifting elements for $d_{ai_1}=d_{ai_2}=40$m and $N_a=N_b=N_e=16$, where no RIS and random phase are used as performance benchmarks. Observing this figure, it is apparent that given the transmit beamforming   proposed  Max-SV or  Max-SLNR, the proposed GPG makes a significant SSR enhancement over no RIS and random phase in terms of SSR. Fixing the RIS phase-shifting method as GPG, the proposed Max-SV outperforms the generalized Max-SLNR when the number of RIS elements is less than 700. Otherwise, there is a converse tendency. Additionally, as the number of RIS elements increases, the SSR performance of the proposed schemes grow gradually. Interestingly, even we close one of two RISs, the performance gain achieved by two-RIS over single-RIS is also attractive.

\begin{figure}[htbp]
\centering
\begin{minipage}[t]{0.45\textwidth}
\centering
\includegraphics[width=0.95\textwidth]{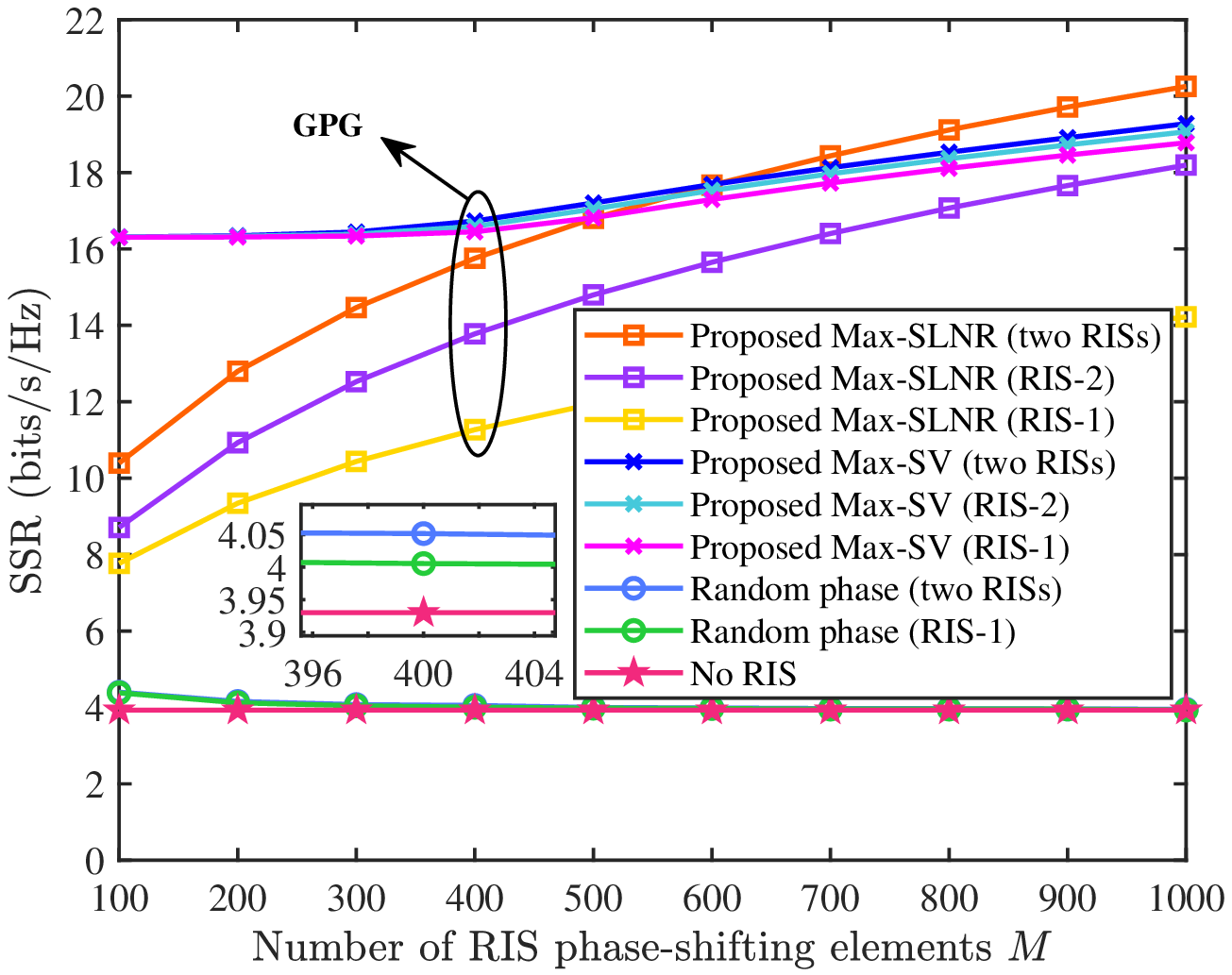}\\
\caption{Curves of SSR versus the number of RIS phase-shifting elements $M$ ($d_{ab}=200$ m).}\label{SSR-200}
\end{minipage}
\begin{minipage}[t]{0.45\textwidth}
\centering
\includegraphics[width=0.95\textwidth]{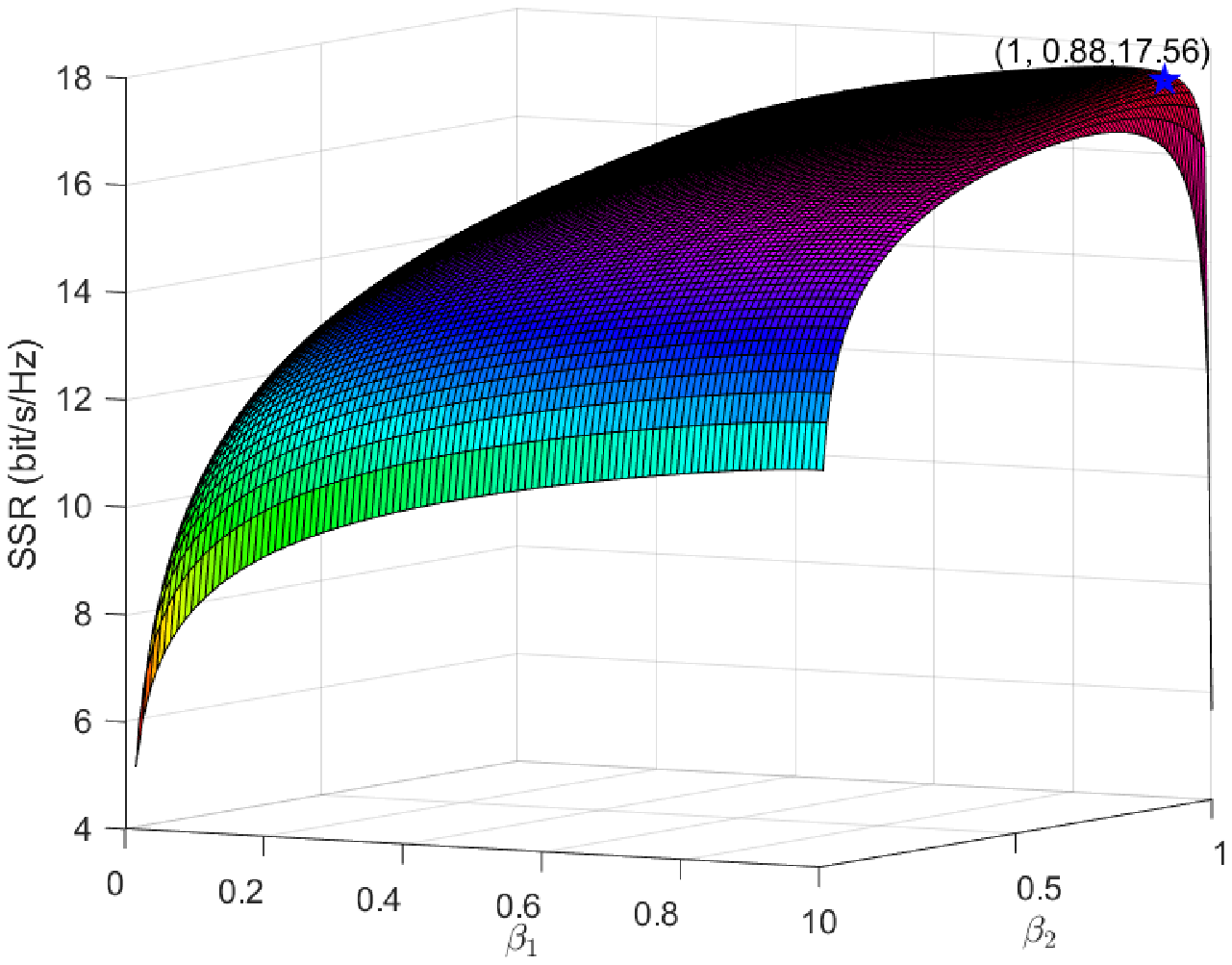}\\
\caption{Curved surface of SSR versus the $\beta_1$ and $\beta_2$ (Max-SV).}\label{PA-SV}
\end{minipage}
\end{figure}

To see the effect of distance on SSR,  Fig.~{\ref{SSR-200}}   plots the curves of SSR versus  the number $M$ of RIS phase-shifting elements by increasing $d_{ab}$ from $70$m to $200$m. Clearly, as $d_{ab}$ increases, the SSR performance of all proposed methods degrades, but there is a similar  performance tendency among those proposed methods.


Fig.~{\ref{PA-SV}} illustrates the curved surface of SSR versus the PA factors $\beta_1$ and $\beta_2$ of the 2D-ES method where the GPG and Max-SV are used for the RIS phase-shifting and transmit beamforming method. As we can seen in the Fig.~{\ref{PA-SV}}, the SSR performance first improves with increasing in PA factors and then decreases dramatically when reaching the optimal point. It seems the optimal values of $\beta_1$ and $\beta_2$  are near one.

\begin{figure}[htbp]
\centering
\begin{minipage}[t]{0.45\textwidth}
\centering
\includegraphics[width=0.95\textwidth]{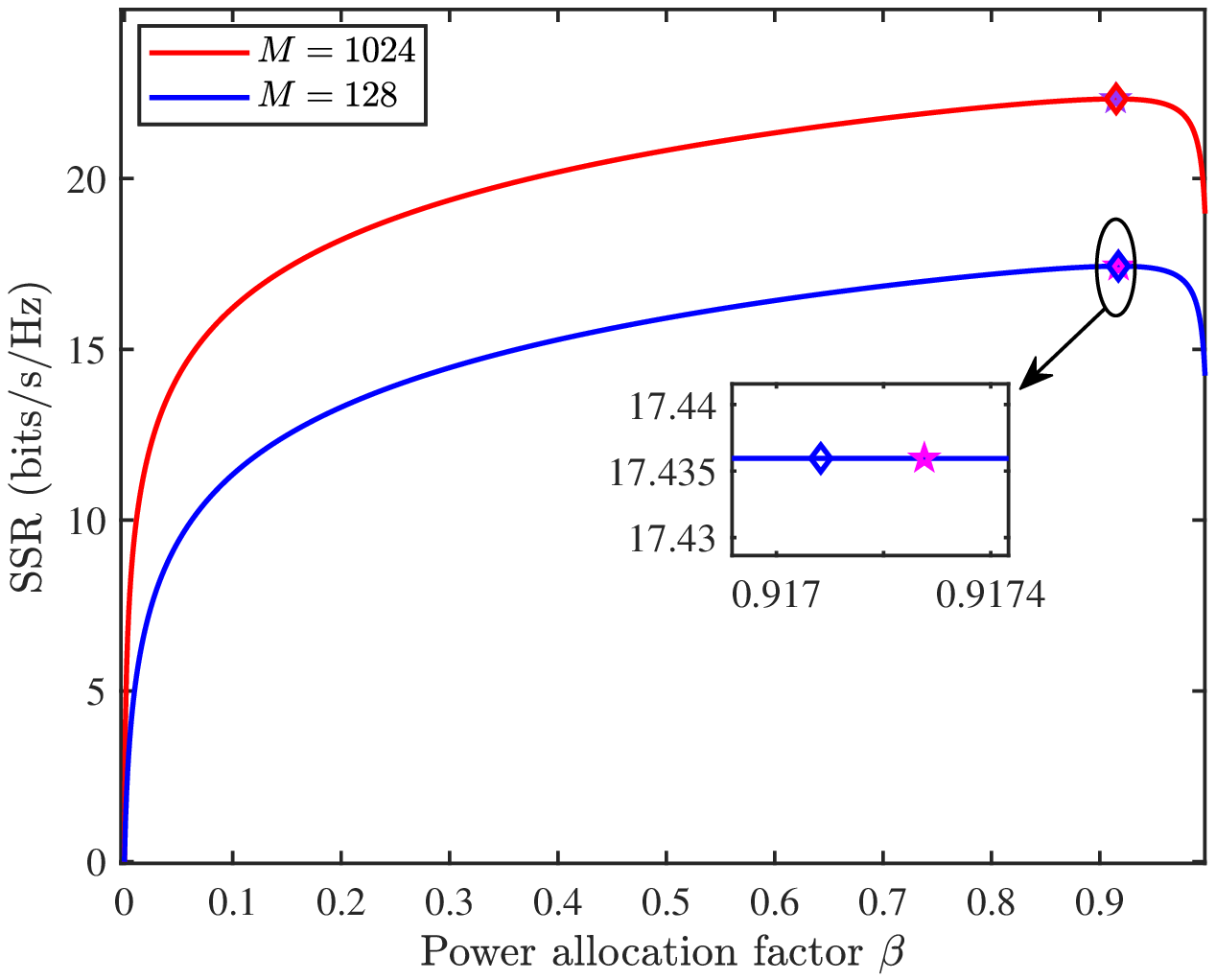}\\
\caption{Curves of SSR versus the PA factor $\beta$ (Max-SV).}\label{PA-LK-SV}
\end{minipage}
\begin{minipage}[t]{0.45\textwidth}
\centering
\includegraphics[width=0.95\textwidth]{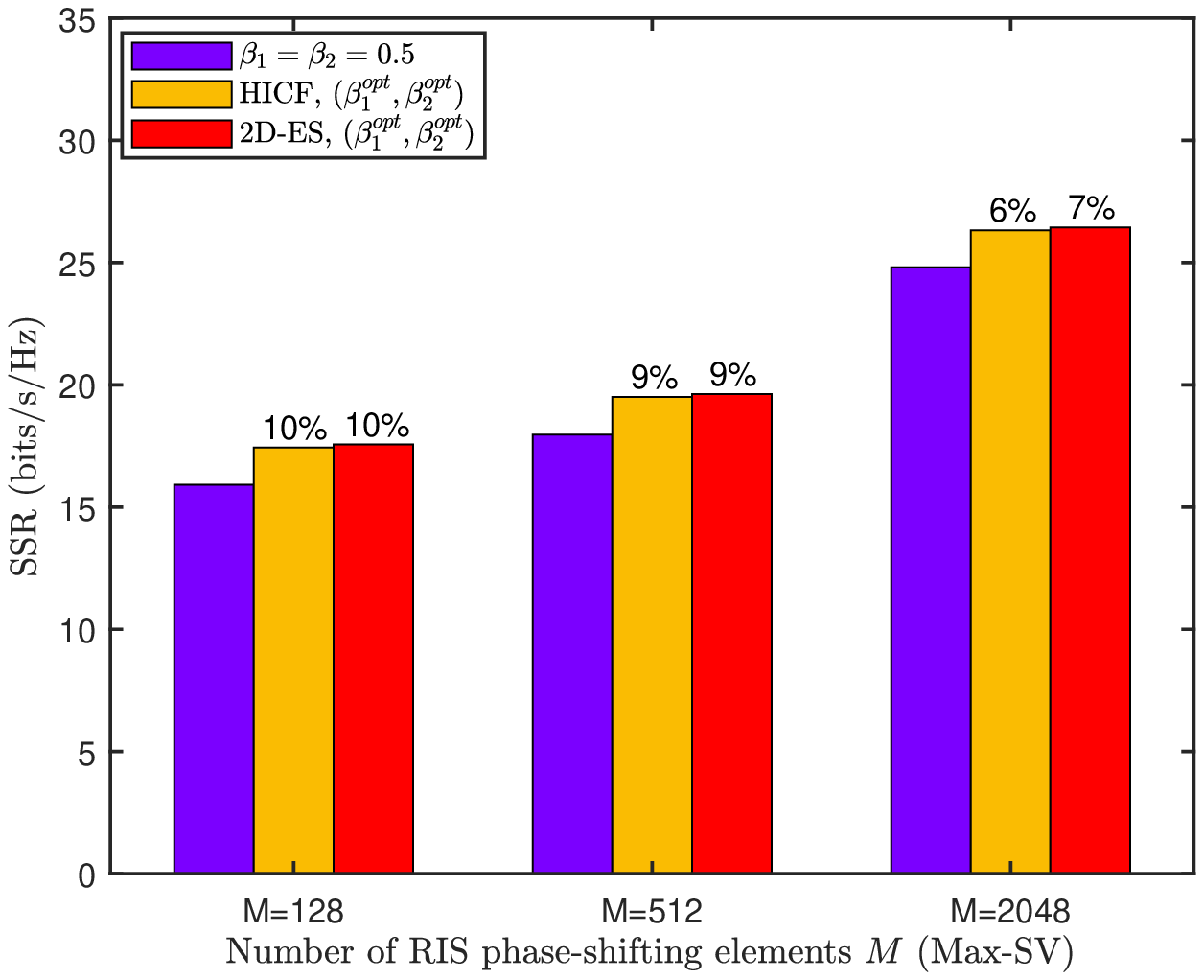}\\
\caption{Histograms of SSR versus the number of RIS phase-shifting elements $M$ (Max-SV).}\label{PA-bar-SV}
\end{minipage}
\end{figure}

 Fig.~{\ref{PA-LK-SV}} depicts the curves of the SSR versus the PA factor $\beta$ for Max-SV method,  and the equal PA is used as a benchmark. It can be seen that 1D-ES and HICF have approximate SSRs for both cases of $M=$128 and 1024. In particularly, observing this figure, we also find the fact that  the SSR  is a concave  function of  $\beta$. In other words, there is one unique extremum in the interval [0,1].


Fig.~{\ref{PA-bar-SV}}  depicts the histograms of the SSR of the proposed HICF versus the number of RIS phase-shifting elements $M$ for Max-SV method with 2D-ES and EPA   as performance benchmarks. At $M=$128, the proposed HICF and 2D-ES can achieve up to ten percent performance gain over EPA. As the number of RIS phase-shifting elements varies from 128 to 2048, the gain shows a slight reduction accordingly.

\section{Conclusion}
In this paper, we have made an investigation of transmit beamforming and PA for a double-RIS-aided two-way DM system. With the help of two RISs, useful controllable multipaths between Alice and Bob can be established. First, the RIS phase-shifting was designed by the GPG criterion. Then the Max-SV transmit beamforming method was proposed, and the Max-SLNR transmit beamforming method is generalized. Finally, a HICF PA algorithm is proposed to enhance the SSR performance with a reduced computational complexity compared with 1D-ES and 2D-ES. From simulation, we can find that the proposed Max-SV and generalized leakage methods approximately triple the SSRs of random phase and no RIS.  Furthermore, the proposed HICF method can provide about a $10\%$ SSR gain over EPA and achieve the same value of SSR as 2D-ES and 1D-ES with a significant reduction in computational complexity. The proposed system and methods may be applied to the future wireless networks like marine communications, UAV network, satellite communications, even 6G.

\section*{Appendix A}
\section*{Proof of (\ref{bata^6})}
In Appendix, we will show how to derive (\ref{bata^6}) from  (\ref{R0-1D-ES}).  By a further simplification, the objective function of (\ref{R0-1D-ES}) can be rewritten in a new simple form
\begin{align}\label{R_beta_1}
R(\beta)=\text{log}_2\frac{q_1\beta^4+q_2\beta^3+q_3\beta^2+q_4\beta+q_5}
{q_6\beta^4+q_7\beta^3+q_8\beta^2+q_9\beta+q_{10}},
\end{align}
where
\begin{align}
q_1&=(s_1-s_2)(s_3-s_4)(-s_7-s_8)^2, \nonumber\\
q_2&=2(s_1-s_2)(s_3-s_4)(-s_7-s_8)(s_7+s_8+\sigma^2_e)+
\big[(s_1-s_2)(s_4+\sigma^2_b)+(s_3-s_4)(s_2+\sigma^2_a)\big]\cdot\nonumber
\end{align}
\begin{align}
&~~~(-s_7-s_8)^2, \nonumber\\
q_3&=(s_1-s_2)(s_3-s_4)(s_7+s_8+\sigma^2_e)^2+2(-s_7-s_8)(s_7+s_8+\sigma^2_e)\big[(s_1-s_2)(s_4+\sigma^2_b)+\nonumber\\
&~~~(s_3-s_4)(s_2+\sigma^2_a)\big]+(-s_7-s_8)^2(s_2+\sigma^2_a)(s_4+\sigma^2_b),\nonumber\\
q_4&=\big[(s_1-s_2)(s_4+\sigma^2_b)+(s_3-s_4)(s_2+\sigma^2_a)\big](s_7+s_8+\sigma^2_e)^2+
2(-s_7-s_8)(s_2+\sigma^2_a)(s_4+\sigma^2_b)\cdot\nonumber\\
&~~~(s_7+s_8+\sigma^2_e),\nonumber\\
q_5&=(s_2+\sigma^2_a)(s_4+\sigma^2_b)(s_7+s_8+\sigma^2_e)^2,\nonumber\\
q_6&=s_2s_4(s_5-s_7-s_8)(s_6-s_7-s_8),\nonumber\\
q_7&=s_2s_4(s_5+s_6-2s_7-2s_8)(s_7+s_8+\sigma^2_e)+(s_5-s_7-s_8)(s_6-s_7-s_8)\big[-s_2(s_4+\sigma^2_b)-\nonumber\\
&~~~s_4(s_2+\sigma^2_a)\big],\nonumber\\
q_8&=s_2s_4(s_7+s_8+\sigma^2_e)^2+(s_5+s_6-2s_7-2s_8)
(s_7+s_8+\sigma^2_e)\big[-s_2(s_4+\sigma^2_b)-s_4(s_2+\sigma^2_a)\big]\nonumber\\
&~~~+(s_5-s_7-s_8)(s_6-s_7-s_8)(s_2+\sigma^2_a)(s_4+\sigma^2_b), \nonumber\\
q_9&=[-s_2(s_4+\sigma^2_b)-s_4(s_2+\sigma^2_a)](s_7+s_8+\sigma^2_e)^2+
(s_5+s_6-2s_7-2s_8)(s_7+s_8+\sigma^2_e)\cdot\nonumber\\
&~~~(s_2+\sigma^2_a)(s_4+\sigma^2_b),\nonumber\\
q_{10}&=(s_2+\sigma^2_a)(s_4+\sigma^2_b)(s_7+s_8+\sigma^2_e)^2.
\end{align}
Let us define
\begin{align}
\phi(\beta)=\frac{q_1\beta^4+q_2\beta^3+q_3\beta^2+q_4\beta+q_5}{q_6\beta^4+q_7\beta^3+
q_8\beta^2+q_9\beta+q_{10}},
\end{align}
then taking the derivative of function $R(\beta)$ with respect to $\beta$  and setting it equal zero give
\begin{align}\label{R-beta-1}
R^{'}(\beta)=\frac{\partial R(\beta)}{\partial \beta}=\frac{1}{\text{In}2\cdot \phi(\beta)}\phi^{'}(\beta)=0.
\end{align}
Considering $\phi(\beta)\neq0$, (\ref{R-beta-1}) reduces to
\begin{align}
\phi^{'}(\beta)=0,
\end{align}
which means that
\begin{align}
&(4q_1\beta^3+3q_2\beta^2+2q_3\beta+q_4)(q_6\beta^4+q_7\beta^3+q_8\beta^2+q_9+q_{10})-\nonumber\\
&(q_1\beta^4+q_2\beta^3+q_3\beta^2+q_4\beta+q_5)(4q_6\beta^3+3q_7\beta^2+2q_8\beta+q_9)=0,
\end{align}
which can be further simplified to
\begin{align}\label{beta4}
&(q_1q_7-q_2q_6)\beta^6+(2q_1q_8-2q_3q_6)\beta^5+(3q_1q_9+q_2q_8-q_3q_7-3q_4q_6)\beta^4\nonumber\\
&+(4q_1q_{10}+2q_2q_9-2q_4q_7-4q_5q_6)\beta^3+(3q_2q_{10}+q_3q_9-q_4q_8-3q_5q_7)\beta^2\nonumber\\
&+(2q_3q_{10}-2q_5q_8)\beta+(q_4q_{10}-q_5q_9)=0.
\end{align}
Notice that (\ref{beta4}) is a sixth-order polynomial since $q_1q_7-q_2q_6\neq0$, let us define
\begin{align}
&\alpha_1=(2q_1q_8-2q_3q_6)/(q_1q_7-q_2q_6),
\alpha_2=(3q_1q_9+q_2q_8-q_3q_7-3q_4q_6)/(q_1q_7-q_2q_6),\nonumber\\
&\alpha_3=(4q_1q_{10}+2q_2q_9-2q_4q_7-4q_5q_6)/(q_1q_7-q_2q_6),\nonumber\\
&\alpha_4=(3q_2q_{10}+q_3q_9-q_4q_8-3q_5q_7)/(q_1q_7-q_2q_6),\nonumber\\
&\alpha_5=(2q_3q_{10}-2q_5q_8)/(q_1q_7-q_2q_6),
\alpha_6=(q_4q_{10}-q_5q_9)/(q_1q_7-q_2q_6).
\end{align}
which means
\begin{align}
f(\beta)=\beta^6+\alpha_1\beta^5+\alpha_2\beta^4+\alpha_3\beta^3+\alpha_4\beta^2+\alpha_5\beta+\alpha_6=0.
\end{align}
This completes the derivation of (\ref{bata^6}).
\ifCLASSOPTIONcaptionsoff
  \newpage
\fi

\bibliographystyle{IEEEtran}
\bibliography{IEEEfull,reference}
\end{document}